\documentclass[twocolumn,twocolappendix]{aastex7}
\usepackage{amsmath, mathtools, amsthm, amssymb}
\usepackage{enumitem}
\usepackage[toc,page]{appendix}
\hypersetup{colorlinks=true, allcolors=cyan}
\usepackage{soul}
\graphicspath{ {./} }
\usepackage{xcolor}
\usepackage{nameref}
\usepackage[super]{nth}
\usepackage{booktabs}
\usepackage{bm}

\newcommand{\msol}{\ensuremath{\rm{M_{\odot}}}}
\newcommand{\NC}{\textsc{NewCluster}}
\newcommand{\HM}{\textsc{AdaptaHOP}}

\usepackage[dvipsnames,svgnames]{xcolor}
\definecolor{aquamarine}{rgb}{0.4, 1.0, 0.83}

\sethlcolor{Lavender} 

\newif\ifshowedel
\showedeltrue   
\ifshowedel
  \newcommand\edel[1]{\bgroup\markoverwith{\textcolor{violet}{\rule[0.5ex]{2pt}{0.8pt}}}\ULon{#1}}
\else
  \newcommand\edel[1]{} 
\fi

\newcommand{\figpath}{.}

\shorttitle{Baryon fraction}
\shortauthors{Changjo Seo et al.}

\begin{document}

\title{
Revisiting the Missing Baryon Problem: Baryon Redistribution Across Halo Evolution}
\author[0009-0004-2529-986X]{Changjo Seo}
\email{tjckdwh1016@yonsei.ac.kr}
\affil{Department of Astronomy, Yonsei University, 50 Yonsei-ro, Seodaemun-gu, Seoul 03722, Republic of Korea: yi@yonsei.ac.kr}

\author[0000-0002-1270-4465]{Seyoung Jeon}
\email{seyoung@kasi.re.kr}
\affil{Korea Astronomy and Space Science Institute, 776, Daedeokdae-ro, Yuseong-gu, Daejeon 34055, Republic of Korea}

\author[0000-0002-2873-8598]{Emanuele Contini}
\email{emanuele.contini82@gmail.com}
\affil{Department of Astronomy, Yonsei University, 50 Yonsei-ro, Seodaemun-gu, Seoul 03722, Republic of Korea: yi@yonsei.ac.kr}

\author[0000-0003-0225-6387]{Yohan Dubois}
\email{dubois@iap.fr}
\affil{Institut d’Astrophysique de Paris, 98 bis Boulevard Arago, F-75014 Paris, France}

\author[0000-0003-2285-0332]{Corentin Cadiou}
\email{corentin.cadiou@iap.fr}
\affil{Institut d’Astrophysique de Paris, 98 bis Boulevard Arago, F-75014 Paris, France}

\author[0000-0001-9939-713X]{San Han}
\email{san.han@iap.fr}
\affil{Institut d’Astrophysique de Paris, 98 bis Boulevard Arago, F-75014 Paris, France}

\author[0000-0002-0858-5264]{J. K. Jang}
\email{starbrown816@yonsei.ac.kr}
\affil{Department of Astronomy, Yonsei University, 50 Yonsei-ro, Seodaemun-gu, Seoul 03722, Republic of Korea: yi@yonsei.ac.kr}

\author[0000-0001-6902-2898]{S. Peirani}
\email{speirani@oca.eu}
\affil{ILANCE, CNRS – University of Tokyo International Research Laboratory, Kashiwa, Chiba 277-8582, Japan}
\affil{Kavli IPMU (WPI), UTIAS, The University of Tokyo, Kashiwa, Chiba 277-8583, Japan}
\affil{Universit\'e C\^ote d'Azur, Observatoire de la C\^ote d'Azur, CNRS, Laboratoire Lagrange, Bd de l'Observatoire,
 CS 34229, 06304 Nice Cedex 4, France\\}

\author[0000-0003-0695-6735]{Christophe Pichon}
\email{pichon@iap.fr}
\affil{Institut d’Astrophysique de Paris, 98 bis Boulevard Arago, F-75014 Paris, France}
\affil{Department of Astronomy \& Space Science, Kyung Hee University, 1732 Deogyeong-daero, Yongin-si, Gyeonggi-do 17104, Republic of Korea}

\author[0000-0002-0184-9589]{Jinsu Rhee}
\email{jinsu.rhee@gmail.com}
\affil{Institut d’Astrophysique de Paris, 98 bis Boulevard Arago, F-75014 Paris, France}

\author[0000-0002-4556-2619]{Sukyoung K. Yi}
\email{yi@yonsei.ac.kr}
\affil{Department of Astronomy, Yonsei University, 50 Yonsei-ro, Seodaemun-gu, Seoul 03722, Republic of Korea: yi@yonsei.ac.kr}


\begin{abstract}
The baryon fraction provides a fundamental diagnostic of baryonic matter distribution within dark matter halos and the baryon cycle. While its cosmic value is well constrained, the baryon content within halos on nearly all scales is significantly lower than the cosmic mean; this is called the ``missing baryon problem.''
Using the high spatial, mass, and temporal resolutions of the \NC\ simulation, we investigate the baryon fraction evolution by tracing halo assembly histories from merger trees and gas tracer particles. 
We identify three distinct evolutionary phases: an early growth phase driven by cold accretion in low-mass halos, a suppression phase likely dominated by supernova feedback, and a recovery phase in massive halos enabled by deeper potential wells and hot-mode accretion.
We further find that the redshift dependence of baryon fraction, particularly in low-mass halos, is influenced by the thermal state of the intergalactic medium, with suppression emerging after the surrounding gas becomes persistently hotter than the halo virial temperature.
In addition, the scatter in baryon fraction at fixed halo mass correlates with the large-scale environment, with baryon-rich halos preferentially residing closer to filamentary structures.
Overall, our results highlight that baryon fraction evolution is governed by the interplay between feedback and environment, requiring a time-resolved perspective on the baryon cycle.
\end{abstract}

\section[]{Introduction}\label{sec:1intro}
\begin{figure}[htb!]
    \includegraphics[width=0.46\textwidth]{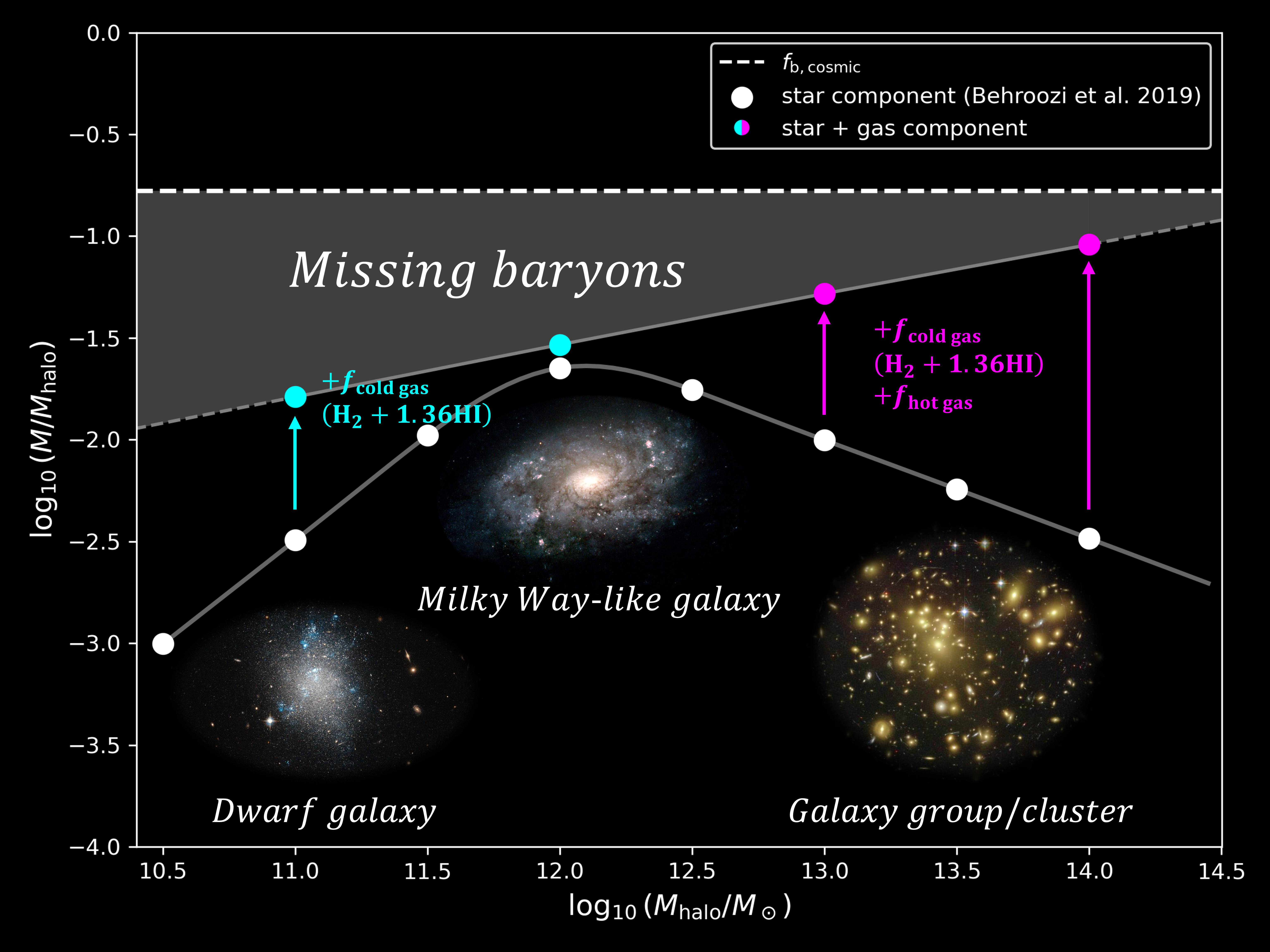}
    \caption{Schematic overview of the missing baryon problem.
White circles show the stellar component from the empirical stellar-to-halo mass relation of \citet{Behroozi2019}. 
Cyan symbols include cold gas, $\mathrm{H_2} + 1.36\,\mathrm{HI}$, using results from \citet{Guo2023} and \citet{Dev2024}, while magenta symbols additionally include the hot gas component from \citet{popesso2026}. The dashed line marks the cosmic baryon fraction, and the shaded region highlights the remaining ``missing baryons.'' }
    \label{fig0}
\end{figure}

The cosmic baryon fraction, defined as the ratio of baryonic to total matter, is an important prediction of any cosmological model
and is tightly constrained by observations of the cosmic microwave background \citep[e.g.,][]{Komatsu2011,planck}. Despite this well-established global value, its connection to the baryonic content of individual dark matter halos remains debatable. 
Figure~\ref{fig0} presents a schematic description of the problem in a rough sense.
Both observations and theoretical studies have consistently shown that the baryon fraction within halos is suppressed relative to the cosmic average, particularly at galaxy and group scales, giving rise to the long-standing “missing baryon problem" \citep[e.g.,][]{Fukugita1998,Cen1999,Bell2003,Hoekstra2005,Bregman2007,Sun2009,McGaugh2010,Shull2012,popesso2025}.

This discrepancy arises from the complex redistribution of baryons across galaxies, the circumgalactic medium (CGM), and the intergalactic medium (IGM), which are dynamically connected. In this cycle, gas accretes from the IGM onto dark matter halos, is processed through the CGM and interstellar medium (ISM), and is subsequently redistributed by star formation and feedback-driven outflows \citep{white1991,Peirani2012,Ford2014,Contini2014,Somerville2015,Peroux2020,Faucher2023,ay2023,Wright2024,Oren2025}.

Supernova (SN) and active galactic nucleus (AGN) feedback processes play a central role in regulating this cycle. SN-driven winds can efficiently remove gas from low-mass halos and enrich CGM and IGM, while AGN feedback can heat and expel gas on larger scales, altering both the thermal state of the surrounding medium and the efficiency of subsequent gas accretion \citep{virial_shock,Oppenheimer2006,Oppenheimer2008,Strickland2009,Puchwein2013,Hopkins2014,Beckmann2017,Rupke2018,Sorini2022,Mitchell2022,Heckman2023}. 

However, observationally probing the CGM and IGM remains highly challenging due to their diffuse nature \citep{werk2014,Tumlinson2017}, and direct constraints on the baryonic content of low-mass halos are still severely limited \citep{Gonzalez2013,Chiu2016,Chiu2018}. As a result, cosmological hydrodynamical simulations provide valuable tools for studying the baryon cycle \citep{Peirani2012}.

State-of-the-art simulations such as 
TNG \citep{TNG2018}, 
EAGLE \citep{EAGLE2015a}, SIMBA \citep{SIMBA2019}, and \textsc{Horizon-AGN} \citep{Dubois2014bHAGN} successfully reproduce many key galaxy observables, including the galaxy stellar mass function, size–mass relation, and cosmic star formation history. However, despite this success at galaxy scales, these simulations often predict significantly different baryon distributions in the CGM and IGM, reflecting the strong dependence on subgrid prescriptions for baryonic physics \citep{Somerville2015,Vogelsberger2020,Crain2023,Iyer2025,rey2025}.

A widely adopted approach to characterizing these differences is to examine the baryon fraction as a function of halo mass at fixed redshift. These distributions provide a useful statistical description of baryon content and have been extensively used to compare simulations and observations \citep{Mitchell2022,ay2023,Wright2024,Ema2025}. However, such population-level statistics do not unambiguously encode the evolutionary pathways of individual halos. 

This limitation arises because the baryon cycle is inherently time-dependent and involves following physical processes which need to be modeled with high enough physical resolutions. Gas accretion, cooling, heating, and ejection operate on different timescales and depend sensitively on halo mass, redshift, and environment \citep{Angle2017,Pandya2021,Mitchell2022}. Since the various processes governing baryon cycling (e.g., accretion, cooling, and feedback) operate on different timescales and do not establish a quasi-steady equilibrium, interpreting the baryon fraction distribution alone is not enough to reveal the underlying physical processes.

To overcome these limitations, an approach of
fine time resolution 
is required, in which the assembly histories of individual halos are explicitly tracked and connected to the physical processes governing baryon accretion and loss. 
In this work, we 
use the \NC\ cosmological zoom-in simulation \citep{newcluster}, which provides high mass, spatial, and temporal resolutions. By using merger trees and gas tracer particles, we can follow the motion of baryons across different phases and environments, enabling a detailed reconstruction of baryon fraction evolution along individual halo histories.

The primary goals of this study are twofold. First, we demonstrate that the baryon fraction distribution at fixed redshift, as usually exercised by observational studies, does not reveal the whole evolutionary pathways of halos and identify the physical origin of this discrepancy. Second, we characterize the evolution of the baryon fraction of individual halos and identify the dominant physical processes shaping its behavior across different mass regimes.

This paper is organized as follows. In Section~\ref{sec:2methods}, we describe the simulation and methodology. In Section~\ref{sec:3result}, we compare the baryon fraction
distribution at fixed redshift with individual evolutionary tracks and investigate their physical origin. In Section~\ref{sec:4discussion}, we explore the connection between baryon fraction scatter and large-scale environment, compare our results with observational constraints, and discuss the limitations of our analysis. Finally, in Section~\ref{sec:6conclusion}, we summarize our main conclusions. 


\section{Methods}\label{sec:2methods}
In this section, we provide a detailed description of the simulation.
Since halo detection and the mechanism of tracer particles are key components of this study, we discuss them in further detail below.
\subsection{Numerical simulations}\label{sec:2method_1sim}
%

\begin{figure*}[htb!]
    \centering
    \includegraphics[width=0.95\textwidth]{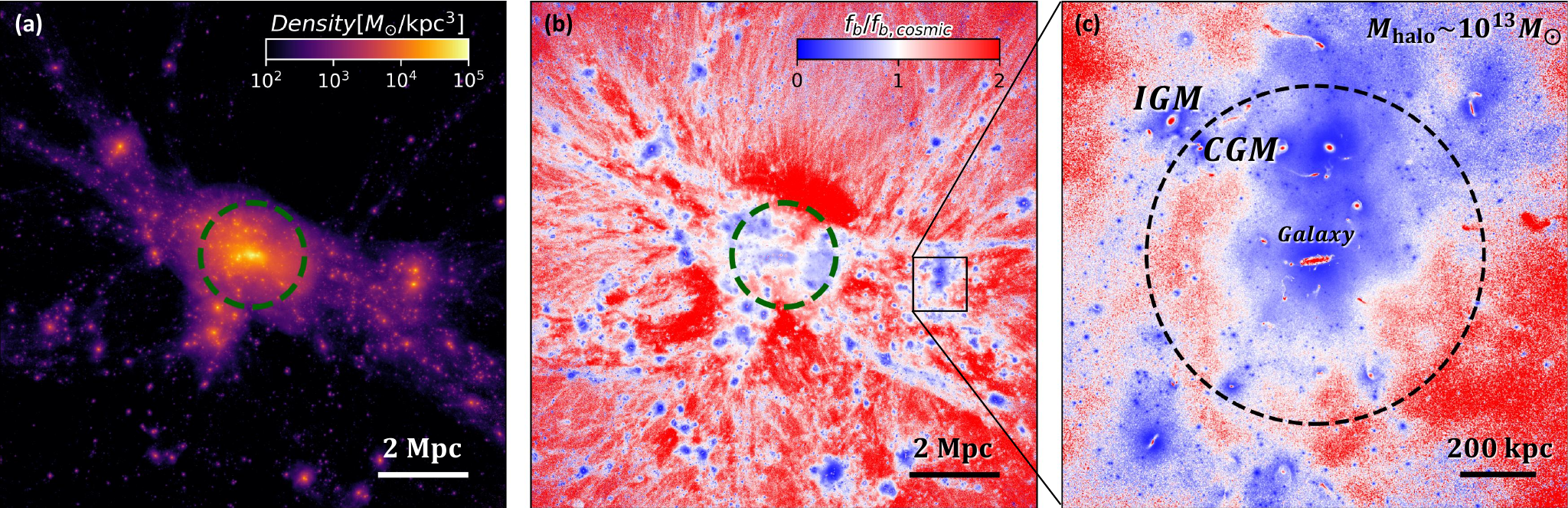}
    \caption{
         Overview of the \NC\ simulation highlighting the large-scale structure and baryon fraction distribution. Panel~(a) shows a dark matter density map, centered on the main cluster of the simulation.
         Panel~(b) shows baryon fraction distribution in the same region as panel~(a). The color-scale represents baryon fraction divided by cosmic baryon fraction which is 0.167 in WMAP7 cosmology.
         Panel~(c) provides a zoomed-in view of an example halo shown in panel~(b), with a halo mass of $\sim 10^{13}\,\msol$.
         The green and black dashed circles represent the virial radii of the main cluster and the target halo, respectively.}
    \label{fig01}
\end{figure*}
%

We use the \NC\ simulation \citep{newcluster}, which is a cosmological zoom-in simulation targeting a cluster-scale region.
At redshift $z = 0$, the virial mass of the main galaxy cluster is $\sim 10^{14.67}\,\msol$, corresponding to that of a Virgo-like cluster.
The simulation was carried out with \textsc{RAMSES}-y\textsc{OMP} \citep{Han2025a_yomp}, which implements an OpenMP-parallelized extension of the \textsc{RAMSES} code originally developed by \citet{Teyssier2002}.
The simulation adopts the \citet{Chabrier2003} initial mass function and is initialized using the WMAP7 cosmology \citep{Komatsu2011}.
The total comoving volume is $(142\,{\rm Mpc})^{3}$ and a zoom-in cluster region at the center with a size of $(28.5\,{\rm Mpc})^{3}$.
In the zoom-in region, the best spatial resolution of gas cells is 68\,pc.
The mass resolutions of stellar and dark matter particles are $2 \times 10^{4}\,\msol$ and $1.3 \times 10^{6}\,\msol$, respectively. 
The corresponding pseudo-lagrangian gas mass resolution is $\sim  10^{5}\,\msol$.

In \NC, radiative cooling and heating are implemented under the assumption of collisional ionization equilibrium.
Cooling rates are evaluated using the tabulated values of \citet{Sutherland1993}, allowing primordial gas to cool down to temperatures of $10^{4}$\,K.
At lower temperatures, metal-line cooling further permits the gas to reach temperatures as low as 0.1\,K~\citep{Dalgarno1972}.
For the reionization model, a uniform ultraviolet background is introduced at $z = 10$, following the model of \citet{Haardt1996}.

Star formation is permitted in individual cells once the gas density exceeds $5\,{\rm H\,cm^{-3}}$, following the gravo-turbulent model of \citet{Kimm2017}. \NC\ does not apply the turbulent Jeans length criterion for star formation, thereby preserving scale invariance as described in \citet{Dubois2021}. The star formation efficiency is calculated following the equation of \citet{Federrath2012}.

Stellar feedback in the simulation is modeled through supernovae (SNe) and stellar winds.
SN feedback is treated by distinguishing between type II and type Ia events (SNII and SNIa), both of which are implemented using the mechanical feedback framework developed by \citet{Kimm2014}. 
This scheme incorporates an empirical momentum boost that depends on the local gas density at the site of the explosion \citep{Kimm2017}. 
Each star particle undergoes SNII explosions after reaching an age of 5\,Myr, and the corresponding feedback energy injected per unit stellar mass is fixed to $e_{\rm SNII} = 0.045$ $\times 10^{51}\,\mathrm{erg}\,\msol^{-1}$.


Massive black holes (MBHs) are seeded in dense, evolved stellar environments based on gas and stellar properties. Seeds form in cells with gas density $n_{\rm cell} > 100\,\rm H\,cm^{-3}$ and stellar mass $\rm m_{*,\rm cell} > 1.88 \times 10^7\,\msol$, with additional constraints that the stellar velocity dispersion is $\sigma_* > 50\,{\rm km\ s^{-1}}$ and stellar age $t_* > 50\,{\rm Myr}$ to avoid short-lived clumps as the sites of black hole seeds. An exclusion radius of $4\,{\rm kpc}$ is adopted to prevent multiple seeds within a single galaxy, ensuring that MBHs form in stable, long-lived stellar systems.

Feedback from AGN is incorporated via a dual-mode model, distinguishing between radio/jet and quasar/heating modes depending on the accretion rate and on the spin of the MBH \citep{Dubois2014a,Dubois2021}. 
The Eddington ratio is defined as $\dot{M}_{\rm BHL} /\dot{M}_{\rm Edd}$, where $\dot{M}_{\rm BHL}$ is the Bondi--Hoyle--Lyttleton accretion rate and $ \dot{M}_{\rm Edd}$ is the Eddington accretion rate. 
When this ratio exceeds 0.01, the AGN operates in the quasar mode, in which isotropic thermal energy is injected to heat the surrounding gas cells, thereby suppressing star formation in the central region.
For values below 0.01, the jet mode is activated, where momentum is imparted to the gas in the direction parallel to the spin axis of the MBH, driving outflows from the host galaxy. 
For theoretical consistency, the Eddington ratio is restricted to values not exceeding unity\footnote{Note, however, that recent studies found strong evidence of super-Eddington accretion of black holes at high redshifts \citep[e.g.,][]{Suh2025}}. Further details of the numerical implementation are provided in the overview paper by \citet{newcluster}.

Since \NC\ has a high stellar mass resolution of $2 \times 10^{4}\,\msol$ and a spatial resolution of $68\,\mathrm{pc}$ at $z=0$, feedback processes can be more effectively resolved even on a cosmological scale, which is expected to provide an advantage for measuring baryon fraction. In addition, \NC\ outputs snapshots at intervals of 15 Myr, which enables detailed tracking of halo assembly histories and baryon fraction evolution across cosmic time. This is particularly important at high redshift, where low-mass halos evolve rapidly and thus require fine temporal resolution to capture their detailed evolutionary behavior.

In Figure~\ref{fig01}, we present an outline of \NC\ at $z=0.44$.
Panel~(a) represents the dark matter density map centered on the main cluster halo in \NC.
Since \NC\ is a zoom-in simulation targeting a cluster region, its dark matter distribution exhibits filamentary structures connected to the central halo. 
Furthermore, since cluster regions are very dense, environmental effects are not confined to cluster subhalos but can extend to surrounding halos beyond the cluster virial radius, which may therefore exhibit properties different from those of field counterparts \citep{Fujita2004, Ragone2007, Han2018, Lacerna2022}. Also, gas expelled by strong cluster-scale outflows can heat and redistribute the surrounding medium, potentially affecting the baryon content of nearby halos. Although \NC\ is cluster-centered, the analyzed main-halo sample extends out to $\sim10\,R_{\rm vir}$ from the main cluster at $z\sim0.44$ and are therefore not restricted to the virialized cluster environment, while still sampling an overdense region.

Panel~(b) shows the baryon fraction distribution in the same region as panel~(a).
In the main galaxy cluster region (green dashed circle), the baryon fraction is generally lower than the cosmic baryon fraction, as indicated by the blue colors, while the brightest cluster galaxy (BCG) itself exhibits a locally enhanced baryon fraction.
In contrast, in the outer regions of the cluster, the baryon fraction exceeds the cosmic value (red), primarily due to gas redistribution driven by virial shocks and large-scale outflows from the main cluster \citep{virial_shock, Biffi2018}.
Panel~(c) presents an example halo, where the baryon fraction exceeds the cosmic value within the galaxy, decreases in the CGM, and increases again in the IGM, reflecting the redistribution of baryons by feedback-driven ejecta from the central galaxy \citep{Cen1999, Shull2012, Sanderson2013, Haider2016, rupke2019,ay2021, Angelinelli2022}.


\subsection{Halo (galaxy) detection and sample selection}\label{sec:2method_2halo}
Galaxies and halos are independently identified with the \HM\ algorithm \citep{Aubert2004, Tweed2009}, which locates density maxima and saddle points in the particle distribution to define preliminary halo centers and extents.
The hierarchical organization of substructures is constructed using the most massive sub-node approach, in which structures connected to the most massive leaf structure are designated as the central component of a host halo, while all other embedded structures are classified as subhalos 
\citep[][]{Tweed2009}.
We set a minimum of 100 particles for structure identification, corresponding to mass thresholds of approximately $10^{6}\,\msol$ for galaxies and $10^{8}\,\msol$ for halos, respectively. With these criteria, we identify 4,452 galaxies and 64,471 halos at $z=0.44$ 
\citep[see][for more details]{jeon2026}.

Since this study focuses on the baryon fraction of main halos 
to exclude orbital motion-associated environmental effects from main halos \citep{Smith2016, Rhee2017, Contini2020, jeon2026}, we exclude subhalos from our sample, yielding a total of 24,227 halos.
The detailed analysis was performed for halos with masses larger than $2 \times 10^{10}\,\msol$, ensuring that each halo is resolved with at least $\sim 10^{4}$ dark matter particles and thus provides numerically robust results; this selection yields a final sample of 361 halos.


%

%

\subsection{Tracer particles}\label{sec:2method_3tracer}

\NC\ adopts an adaptive mesh refinement (AMR) approach, which follows an Eulerian framework in which the gas dynamics is computed on a grid-based representation of the fluid.
AMR simulations allow high spatial resolution to be dynamically concentrated in regions of interest following the gravitational collapse, and multi-scale gas dynamics across a wide dynamic range, with an efficient shock-capturing method \citep{Shea2005, Hubber2013}.
Although this approach provides computational efficiency and offers advantages in specific aspects, it has an inherent limitation: due to its grid-based, Eulerian nature, it is difficult to track the Lagrangian history of gas elements.
This limitation complicates the tracking of gas flows and, in particular, makes it difficult to accurately quantify inflow and outflow masses separately as well as the gas accretion history. 

To overcome these disadvantages, \NC\ adopts tracer particles based on Monte-Carlo approach which follow the gas distribution of the simulation box \citep{Genel2013, Cadiou2019}.
Each tracer particle moves probabilistically in order to match the mass exchange processes in the simulation, including inter-cell gas fluxes, star formation, MBH accretion, and mass ejection from feedback processes.
Depending on the underlying physical process, tracers can transition back and forth between gas, star, and MBH tracer states.
This approach enables the reconstruction of Lagrangian histories of baryonic matter throughout the simulation \citep[e.g.,][]{Cadiou2022,Kocjan2024}.
\NC\ includes a total of 267,997,944 tracer particles, corresponding to a uniform tracer mass resolution of $5.17 \times 10^{5}\,\msol$ per tracer.
In \NC, although the spatial distribution of gas tracer particles closely follows that of the gas, their finite number and relatively poor mass resolution compared to the gas (they are five times more massive than the pseudo gas mass resolution) can introduce minor discrepancies. For more detailed descriptions of the implementation and its validation, we refer the reader to \cite{Cadiou2019} (see also \citealp{newcluster}).


\begin{figure*}[htb!]
    \centering
    \includegraphics[width=0.95\textwidth]{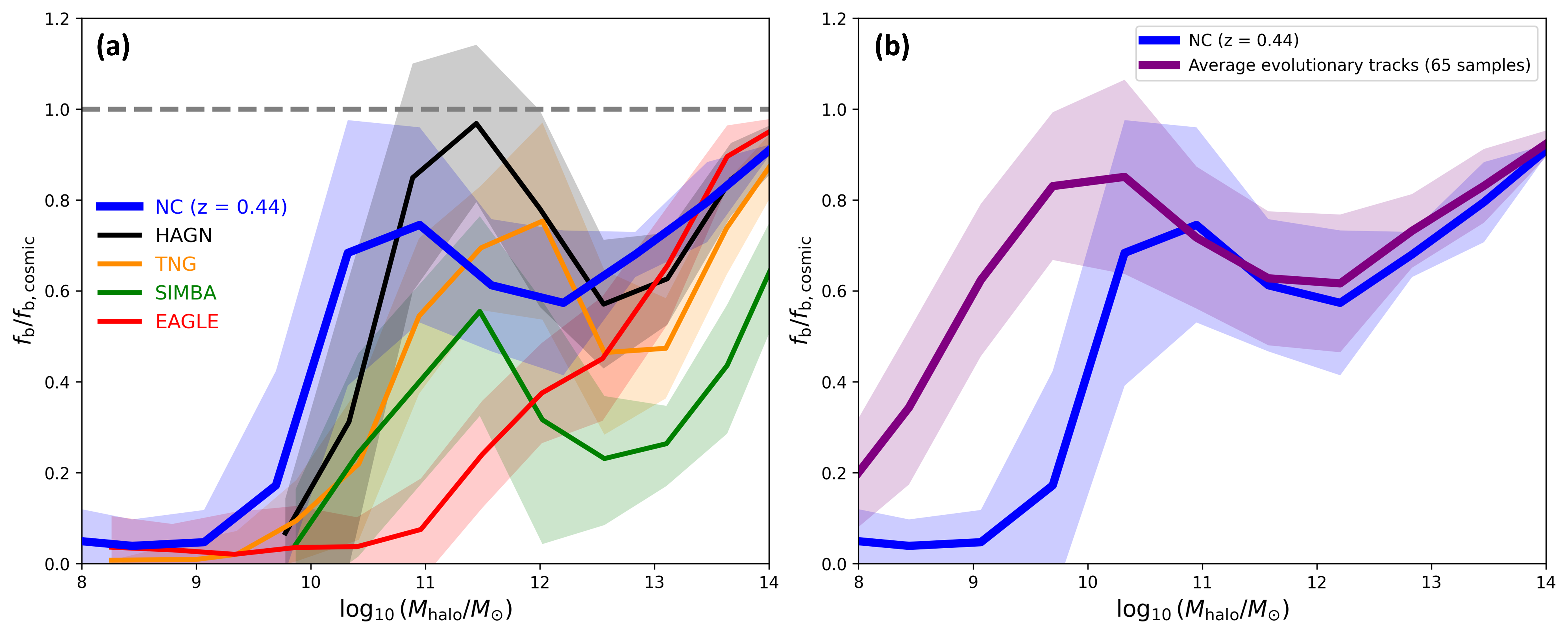}
    \caption{
        Comparison of baryon fraction as a function of halo mass across simulations and their evolutionary tracks in \NC. Panel~(a) shows the mean baryon fraction normalized by the cosmic baryon fraction, $f_{\rm b}/f_{\rm b,cosmic}$, as a function of halo mass for different cosmological simulations. The blue line shows the result from \NC\ (NC) at $z = 0.44$, while the black, orange, green, and red lines correspond to \textsc{Horizon-AGN} (HAGN), TNG, SIMBA, and EAGLE at $z = 0$, respectively. The results for TNG, SIMBA, and EAGLE are taken from \citet{ay2023}. Shaded regions indicate the mean baryon fraction with $\pm1\sigma$ halo-to-halo scatter. The horizontal dashed line marks the cosmic baryon fraction. Panel~(b) shows the evolutionary tracks of baryon fraction of the massive halos and $f_{\rm b}$ distribution in \NC. The purple line represents the mean evolutionary tracks constructed from 65 halos with halo masses $M_{\rm halo} > 10^{11}\,\msol$ at $z = 0.44$, with the shaded region indicating the $\pm1\sigma$ halo-to-halo scatter, while the blue line shows the baryon fraction–halo mass relation at $z = 0.44$, identical to that in panel~(a).}
    \label{fig02}
\end{figure*}

\section{Results}\label{sec:3result}
In this section, we compare the baryon fraction distribution with individual evolutionary tracks.
Since the shape of the evolutionary tracks is a key element of this analysis, its physical origin is discussed in detail.
For convenience, we hereafter denote the baryon fraction as $f_{\rm b}$.
\subsection{Baryon fraction distribution vs evolutionary tracks}\label{sec:3result:1distri}
\begin{deluxetable*}{lcccc}
\tablecaption{Summary of the simulation specifications used in Figure~\ref{fig02} (a).\label{tab:simulations}}
\tablehead{
\colhead{Simulation} &
\colhead{Box size [cMpc]} &
\colhead{Dark matter mass resolution [$\msol$]} &
\colhead{Stellar mass resolution [$\msol$]} &
\colhead{Code}
}
\startdata
TNG100    &110.7 & $7.5 \times 10^{6}$ & $1.4 \times 10^{6}$ & \textsc{AREPO} \\
EAGLE          & 100& $9.7 \times 10^{6}$ & $1.8 \times 10^{6}$ &  \textsc{GADGET3} \\
SIMBA        & 147.6  & $9.6 \times 10^{7}$ & $1.8 \times 10^{7}$  & \textsc{GIZMO} \\
\textsc{Horizon-AGN} &142& $8\times10^{7}$& $2 \times 10^{6}$ &\textsc{RAMSES}\\
\NC       &142 (28.5 zoom-in)   & $1.3 \times 10^{6}$ & $2 \times 10^{4}$ &  \textsc{RAMSES} \\
\enddata
\end{deluxetable*}
In this subsection, we compare the distribution of baryon fraction, $f_{\rm b}$, as a function of halo mass at a given redshift with the evolutionary tracks that trace the growth histories of massive halos selected at $z=0.44$. 
To enable this comparison, it is necessary to measure $f_{\rm b}$ for each halo. 
In this paper, we adopt the virial radius provided by the \HM\ output as the halo boundary for \NC\ and Horizon-AGN. We note that the results for TNG, SIMBA, and EAGLE are taken from \citet{ay2023}.
In our analysis, baryon fraction is measured as the total baryonic mass within the adopted halo boundary divided by the halo mass, where the halo mass is the sum of baryonic and dark matter masses.

It should be noted that \NC\ is currently at $z=0.44$ and is still going.
We use the last snapshot data at the time of writing this manuscript, while the other simulations mentioned above are analyzed at the local universe. This redshift range ($z>0.44$) is sufficient for our purpose because it covers most of the halo mass range we are interested in, which will be clear in the following paragraphs. Moreover, as shown in Figure~\ref{figA}, the redshift evolution at $z<0.44$ is small.

In Figure~\ref{fig02}, panel~(a) presents the distribution of the halo baryon fraction, for various cosmological simulations. Their key specifications are summarized in Table~\ref{tab:simulations}. TNG \citep{TNG2018}, EAGLE \citep{EAGLE2015a}, SIMBA \citep{SIMBA2019}, and \textsc{Horizon-AGN} \citep{Dubois2014bHAGN} are full-volume simulations, providing statistically significant samples but with relatively lower resolution. 
In contrast, \NC\ is a zoom-in simulation, which is designed to achieve higher resolution in selected regions at the cost of a smaller sample size, which allows more detailed numerical treatments.
In particular, the median effective gas cell radius in \NC\ is $2.69\,\rm ckpc$, compared to $15.8\,\rm ckpc$ in TNG100, which is also a mesh-based simulation, while the mean effective cell radius of star-forming gas is $85\,\rm cpc$ in \NC\ (as opposed to $355\,\rm cpc$ in TNG100). This substantially-improved spatial resolution makes \NC\ particularly well suited for the present analysis.

In this figure, the baryon fraction of each halo is normalized by the cosmic baryon fraction of each simulation and shown as a function of halo mass.
Considering the mass resolution of each simulation, halos in \NC\ are resolved with at least $10^{4}$ dark matter particles above halo masses of $10^{10}\,\msol$, while the other simulations reach a similar level of numerical robustness only above $10^{11}\,\msol$.
Halos above these limits can therefore be regarded as numerically robust systems.


In all simulations, the baryon fraction approaches the cosmic mean toward the cluster mass scale, reflecting the increased ability of massive halos to retain hot gas within their deep gravitational potential wells. In addition, baryon fraction is also known to be sensitive to subgrid physics prescriptions, most notably feedback processes \citep{ay2023, Wright2024, Ema2025, Iyer2025}. While \NC\ and \textsc{Horizon-AGN} are both run with similar versions of the \textsc{RAMSES} code, they show notable offsets in the first peak both horizontally and vertically.
These differences may reflect a combination of numerical resolution and feedback prescriptions, including the higher resolution and somewhat stronger SN feedback adopted in \NC.

All simulations but EAGLE show a similar overall $N$-shaped trend (rise-fall-rise).
The two most recent simulations, \NC\ and TNG, exhibit similar absolute baryon fraction values and overall trends.
The horizontal offset in the peak position likely reflects differences in numerical resolution and the adopted subgrid physics between the two simulations.
EAGLE does not show the $N$-shaped behavior.
This is likely related to the combined effect of relatively strong SN feedback and weak AGN feedback.
Compared to TNG and SIMBA, the stronger SN feedback in EAGLE suppresses the baryon fraction more efficiently in low-mass halos, while the absence of a jet-mode AGN prescription \citep{EAGLE2015a, EAGLE2015b} makes AGN-driven outflows less effective in massive halos, allowing the baryon fraction to quickly recover in massive halos.
The interplay between these two effects may contribute to the absence of the characteristic $N$-shaped trend \citep{ay2023, Wright2024}.

Overall, these comparisons suggest that variations in the baryon fraction distribution likely reflect a combination of differences in the adopted subgrid physics and numerical resolution.
To further examine whether the distinct distribution in \NC\ is related to the zoom-in simulation setup, we additionally compare other \textsc{RAMSES}-based zoom-in simulations in Appendix~\ref{sec:app}.

Given that most simulations display a characteristic $N$-shaped trend, we next investigate whether the massive halos in \NC\ sample evolve along this trend.
To investigate this, we select relatively massive halos with $M_{\rm halo}>10^{11}\,\msol$ at $z=0.44$ in \NC, resulting in a sample of 65 main halos.
We construct merger trees using our in-house code \textsc{PhantomTree} (see Appendix~A of \citealp{Jeon2025} for details), identify their progenitors, and trace their evolutionary histories, from which individual evolutionary tracks are constructed.

Panel~(b) of Figure~\ref{fig02} shows the mean evolutionary track obtained from these 65 halos (purple), overlaid with the baryon fraction distribution of the \NC\ simulation shown in panel~(a).
In the high-mass regime ($M_{\rm halo}>10^{11}\,\msol$), the evolutionary tracks are almost identical to the distribution.
In contrast, in the low-mass regime ($M_{\rm halo}<10^{11}\,\msol$), the baryon fraction is significantly suppressed.
This indicates that halos did not evolve by simply following the distribution at present-day and that their evolutionary paths differ, particularly at low masses.
A key distinction between the two curves in the low-mass regime lies in their evolutionary context.
Unlike the low-mass halos in the $f_{\rm b}$ distribution at $z=0.44$, the low-mass regime of the evolutionary tracks corresponds to earlier stages through which these halos evolved, and therefore to higher redshifts.
\begin{figure}[tb!]    \includegraphics[width=0.45\textwidth]{\figpath/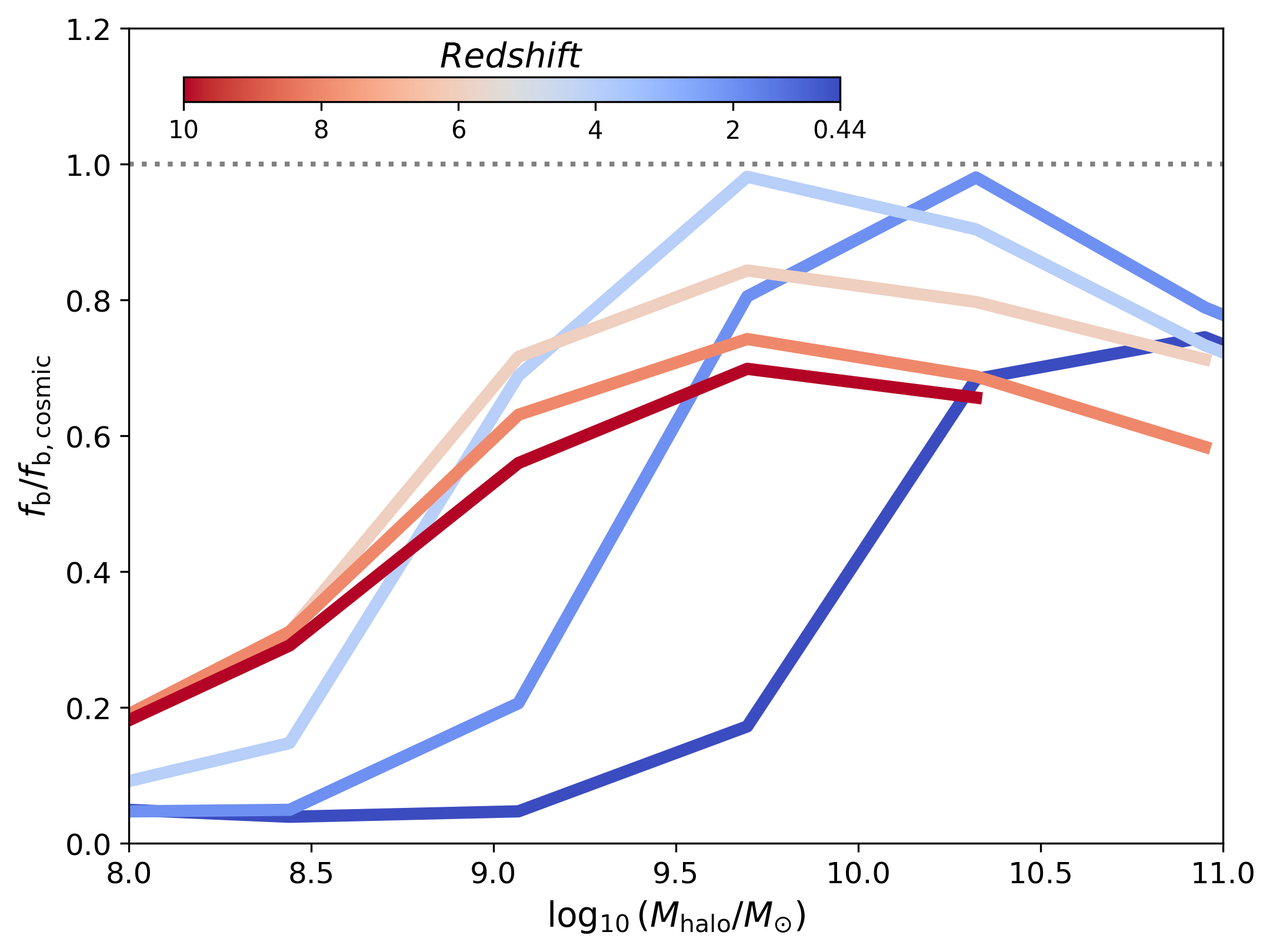}
    \caption{Redshift dependence of $f_{\rm b}$ distribution, normalized by the cosmic baryon fraction. Each colored curve represents the mean $f_{\rm b}/f_{\rm b,cosmic}$ at a given redshift, with the color indicating redshift as shown by the color bar. The horizontal gray dashed line marks the cosmic baryon fraction.}
    \label{fig03}
\end{figure}

We note that these evolutionary tracks are subject to a progenitor-selection effect, as they are constructed by selecting relatively massive halos at $z=0.44$ and tracing their progenitors backward in time. By construction, their low-mass progenitors are therefore selected to undergo substantial subsequent mass growth and are not representative of the general low-mass halo population in terms of their mass assembly histories. However, we find no clear systematic offset in $f_{\rm b}$ between the selected progenitors and the general halo population at comparable halo mass and redshift (plot not shown). Thus, while the sample is biased in terms of mass assembly history, this selection does not appear to introduce a strong systematic bias in the $f_{\rm b}$ evolutionary tracks.

\begin{figure}[htb!]
    \includegraphics[width=0.45\textwidth]{\figpath/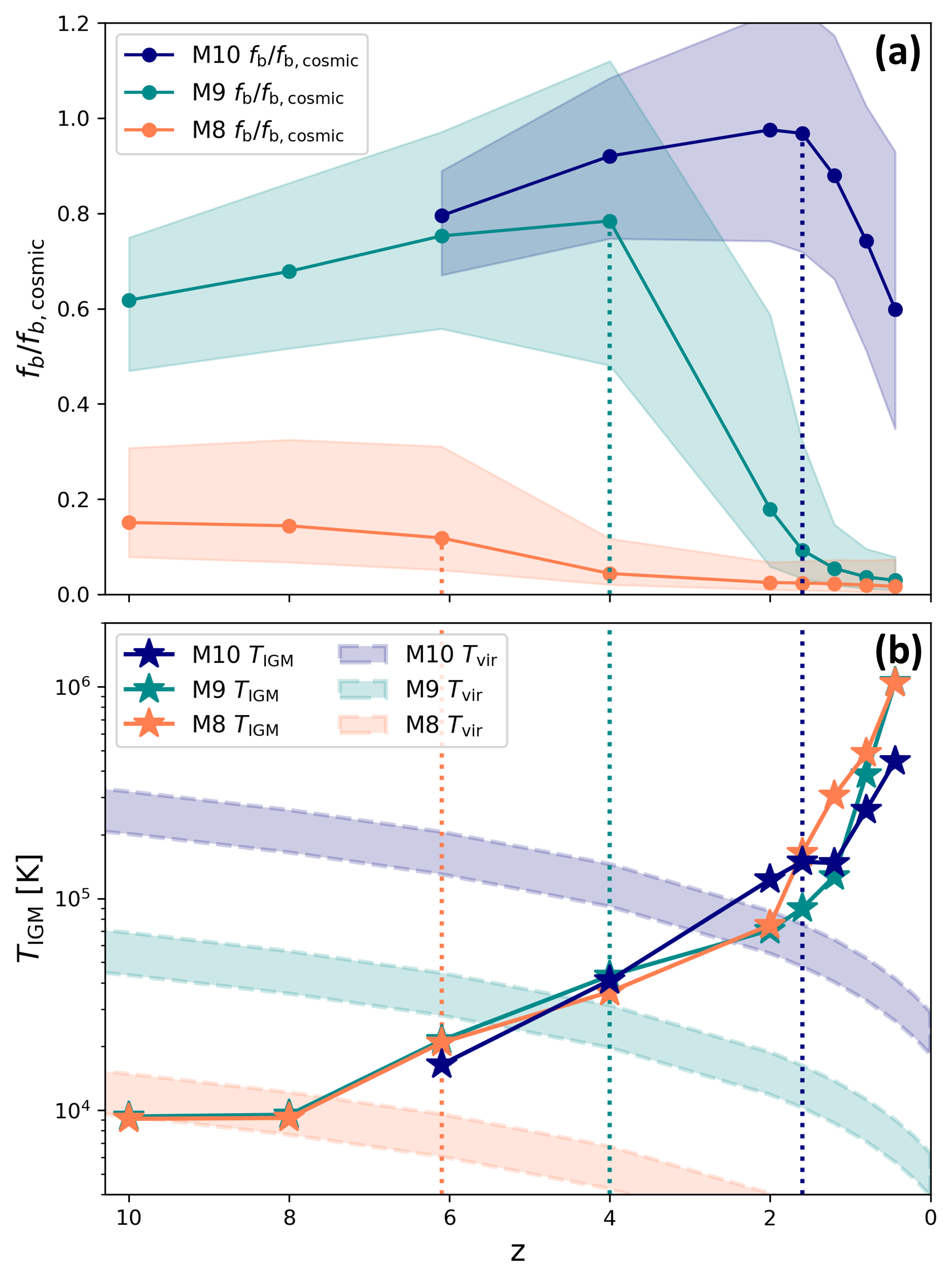}
    \caption{Redshift dependence of baryon fraction and IGM thermal state in low-mass halo samples. Panel~(a) shows three halo samples at each redshift: M8 ($M_{\rm halo}$ = (1--2)$\times10^{8}\,\msol$), M9 ($M_{\rm halo}$ = (1--2)$\times10^{9}\,\msol$), and M10 ($M_{\rm halo}$ = (1--2)$\times10^{10}\,\msol$), selected to probe the low-mass regime. At each redshift, 1000, 400, and 200 halos are randomly selected for the M8, M9, and M10 samples, respectively. Solid lines indicate the median baryon fraction of each sample, normalized by the cosmic baryon fraction, while the shaded regions denote the 16th–84th percentile range. Panel~(b) shows the median gas temperature of the IGM measured within $1$–$2\,R_{\rm vir}$ for each halo sample, plotted as a function of redshift (solid lines). The virial temperatures of the corresponding halo samples are shown as the shaded bands. The dotted lines indicate the redshifts at which the baryon fractions of the M8, M9, and M10 samples begin to be suppressed.}
    \label{fig04}
\end{figure}

In conclusion, the difference between the evolutionary tracks and the $f_{\rm b}$ distribution at $z=0.44$ suggests a redshift dependence in the low-mass regime, indicating that low-mass halos may experience different environments and evolutionary processes across redshift.
To clearly illustrate the redshift dependence in the low-mass regime, we present Figure~\ref{fig03}, which shows $f_{\rm b}$ as a function of halo mass at various redshifts.
In the low-mass regime, baryon fraction suppression appears progressively from lower to higher halo masses as cosmic time advances. This trend indicates that baryon fraction suppression extends toward progressively more massive halos with increasing time.

\subsection{Thermal state of IGM}

In \NC, we note that the redshift dependence of baryon fraction suppression, when examined at fixed halo mass, is found to be primarily driven by a reduction in the gas mass rather than the stellar mass.
This indicates a systematic decline in gas accretion efficiency at fixed halo mass toward lower redshift.

To investigate the redshift dependence of baryon fraction suppression in low-mass halos, we construct three halo samples at fixed halo mass across different redshifts. The M8, M9, and M10 samples correspond to the halos with masses of $M_{\rm halo}=(1\mbox{--}2)\times10^{8}\,\msol$, $M_{\rm halo}=(1\mbox{--}2)\times10^{9}\,\msol$, and $M_{\rm halo}=(1\mbox{--}2)\times10^{10}\,\msol$, respectively, at the current $z$.
The redshift dependence of the median baryon fraction is shown in panel~(a) of Figure~\ref{fig04}. The thermal state of the surrounding IGM is quantified separately in panel~(b), where we present the median IGM temperature measured from gas within $1\mbox{--}2\,R_{\rm vir}$ of each halo.
The shaded bands represent the virial temperatures of halos in each sample, estimated using the standard relation (\citealt{Bryan98}):
\begin{equation}
\begin{split}
T_{\rm vir} =\;& 1.59 \times 10^{3} 
f_{\rm T}\left(\frac{\mu}{0.6}\right)
\left(\frac{M_{\rm halo}}{10^{8} \msol}\right)^{2/3} \\
& \times \left[\frac{\Delta_{\rm c}(z)E^2(z)}{18\pi^2}\right]^{1/3}\, {\rm K}\, ,
\end{split}
\end{equation}
where $f_{\rm T}\simeq 0.8$ is a fudge factor for gas temperature, $\mu$ is the mean molecular weight, $\Delta_{\rm c}(z)$ is the virial overdensity with respect to the critical density of the Universe at redshift $z$, and $E^2(z)\simeq\Omega_{\rm m}(1+z)^3+\Omega_\Lambda$, where $\Omega_{\rm m}$ and $\Omega_\Lambda$ are the cosmic matter density and the cosmic dark energy density, respectively. In this work, we adopt a simplified form of this relation by assuming $\mu = 0.6$, appropriate for ionized primordial gas, and since we focus on the redshift range $z \sim 0.44\mbox{--}10$, the virial temperature reduces to
\begin{equation}
T_{\rm vir} \simeq 8.5 \times 10^{3} \, {\rm K} \,
\left(\frac{M_{\rm halo}}{10^{8} \,\msol}\right)^{2/3}
\left(\frac{1+z}{10}\right).\end{equation}
For the M8 sample, we find that baryon fraction suppression begins at $z\sim6$, whereas the M9 and M10 samples exhibit mild increases in baryon fraction persisting until $z\sim4$ and $z\sim1.6$, respectively. 
The baryon content is regulated by the tug of war between $T_{\rm IGM}$ and $T_{\rm vir}$ with a delayed response associated with the filtering scale. As shown in panel~(b), the onset of strong suppression occurs after the IGM becomes hotter than the halo virial temperature for a sufficient period of time. 
It is progressively more difficult with time for the surrounding IGM gas to be accreted onto these halos beyond this point \citep{Gnedin2000,Rasera2006,Okamoto2008, noh2014}. 
As lower-mass halos have shallower gravitational potential wells, their gas accretion efficiency is more sensitive to the thermal state of the surrounding medium, causing this critical transition to occur at earlier times, and therefore leading to an earlier onset of baryon fraction suppression in the M8 sample.

The close connection between baryon fraction suppression and the thermal state of the surrounding IGM suggests that IGM heating is a key driver of the reduced gas content in low-mass halos. Heating of the IGM in New-Cluster arises primarily from three mechanisms: cosmic reionization, SN feedback, and AGN feedback.

(i) Reionization: since \NC\ does not explicitly include radiative transfer, reionization is implemented using a uniform ultraviolet background model.
In \NC, reionization begins at $z=10$ and is almost completed at $z\simeq6.1$.
During this process, cold gas is heated to temperatures of order $\sim 10^{4}\,\mathrm{K}$, making it more difficult for low-mass halos to accrete it \citep{Van2011,Faucher2011}.

(ii) SN feedback: in \NC, some halos that form early already host star formation at $z\gtrsim10$.
The resulting SN explosions generate shock waves that propagate outward from the central regions, releasing energy into the surrounding medium.
This process not only expels gas from halos but also heats the surrounding gas, including material located outside the halo \citep{Dekel1986,Hopkins2014}.

(iii) AGN feedback: AGN feedback is one of the most powerful feedback mechanisms operating within galaxies \citep{Croton2006,Fabian2012,King2015}.
However, in \NC, due to the stringent black hole seeding criteria, black holes are rarely seeded in halos with masses below $10^{11}\,\msol$.
As a result, AGN activity is rare at $z\gtrsim5$, and AGN feedback is unlikely to contribute significantly to the suppression of the baryon fraction at early times for our samples.
Instead, AGN feedback becomes increasingly important at later epochs, particularly from $z\sim2$ onward, when many halos enter the jet-mode stage.
At these redshifts, AGN-driven outflows substantially heat the surrounding IGM and are expected to suppress the baryon fraction in more massive halos.
 
After the onset of reionization ($z \simeq 10$), the uniform ultraviolet background radiatively heats the gas to temperatures of $\sim 10^{4}\,\mathrm{K}$, which can contribute to the suppression of the baryon fraction. However, as shown in Figure~\ref{fig04}, no immediate baryon fraction suppression is observed in any of the samples. 


For the M8 sample, the IGM temperature exceeds the halo virial temperature around the onset of reionization, while baryon fraction suppression emerges only near the end of reionization. This delayed suppression is consistent with the filtering scale picture and suggests that reionization-driven heating dominates the baryon fraction suppression in M8 halos.
This is consistent with the radiative transfer simulations of \citet{Katz2020}.


Although SN feedback already exists from $z\sim10$ in early-formed galaxies, it does not increase the global IGM temperature above the $\sim 10^{4}$ K level established by reionization by the time of $z\sim6$. For the M9 and M10 samples, however, the regime where $T_{\rm IGM} > T_{\rm vir}$ is reached only after the IGM temperature has risen above the reionization heating level. Since AGN jet-mode feedback is still not dominant during the redshift interval $z\sim6$--$2$, the subsequent baryon fraction suppression in the M9 and M10 halos is most naturally explained by SN feedback, which also appears to be the dominant source of the gradual IGM heating during this epoch.

Toward lower redshift ($z\lesssim2$), the IGM temperature rises sharply, coinciding with the onset of widespread AGN jet-mode activity. This behavior is consistent with previous studies showing that the thermal state of the IGM at low redshift is strongly influenced by AGN feedback \citep[e.g.,][]{Sorini2022}. The associated heating of the surrounding gas is therefore expected to extend baryon fraction suppression to halos more massive than the M9 and M10 samples. Consequently, our results suggest that the dominant source of IGM heating evolves with cosmic time: reionization-driven heating primarily affects the M8 sample, SN feedback appears to dominate the gradual IGM heating during $z\sim6$--$2$, which is responsible for suppression in the M9 and M10 samples, and AGN feedback becomes increasingly important at lower redshift, extending suppression to higher halo masses than those affected by SN feedback alone. Our time-resolved analysis further demonstrates that this thermal evolution of the IGM is directly linked to the progressive suppression of the baryon fraction from low- to high-mass halos.
\subsection{Feedback processes shaping the evolutionary tracks}\label{sec:3result:2evol}

\begin{figure}[htb!]
    \includegraphics[width=0.45\textwidth]{\figpath/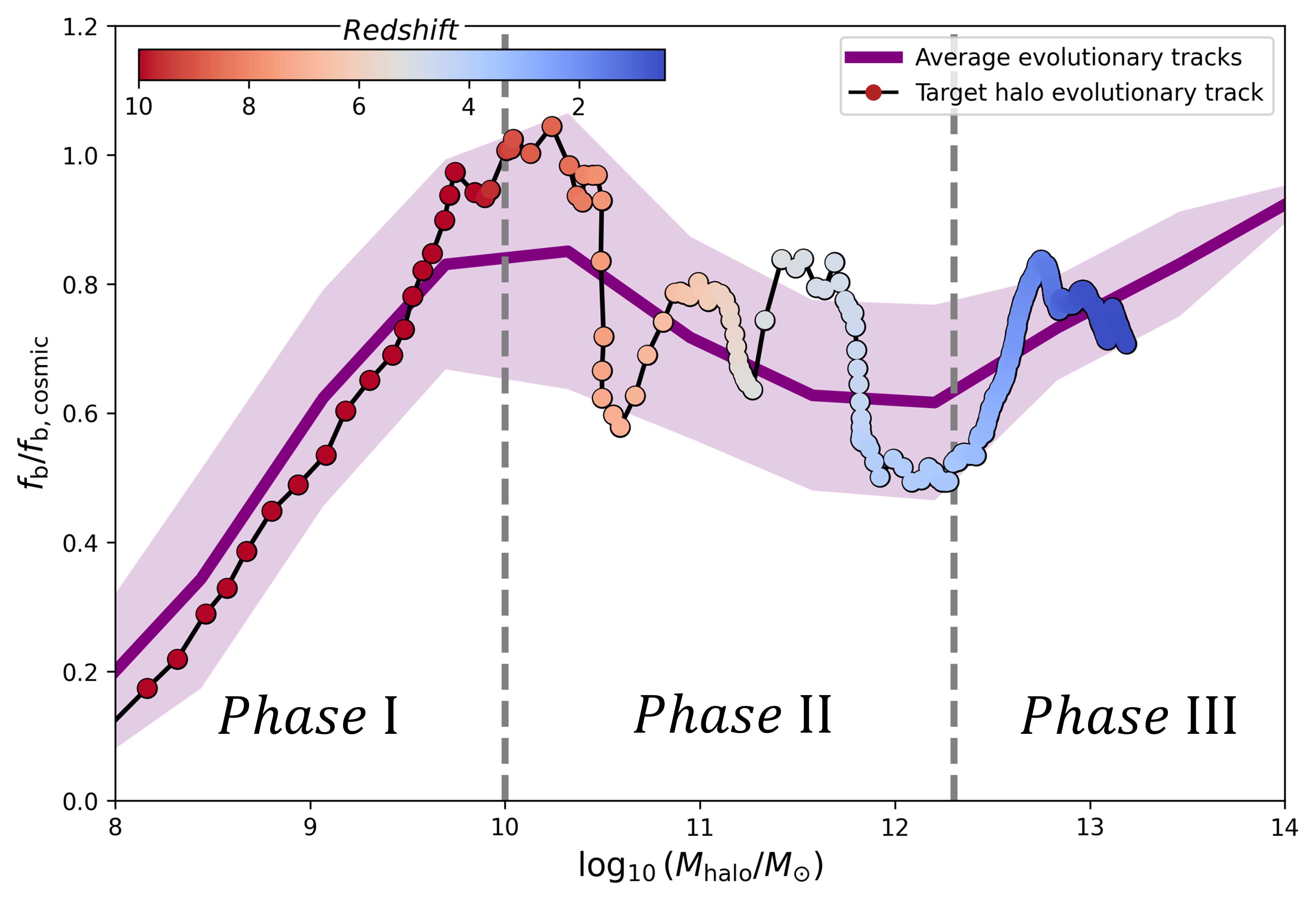}
    \caption{Evolution of the baryon fraction, normalized by the cosmic value, as a function of halo mass.
    The purple curve shows the mean evolutionary track of the halo population ($M_{\rm halo} > 10^{11}\,\msol$ at $z = 0.44$), while the shaded region represents the $\pm1\sigma$ halo-to-halo scatter.
    The black curve indicates the evolutionary track of the selected target halo, while colored dots denote individual snapshots, color-coded by redshift.
    The two vertical gray dashed lines mark the halo masses corresponding to the peak and trough in the baryon fraction evolution, respectively.}
    \label{fig05}
\end{figure}
In this subsection, we discuss which feedback processes primarily shape the overall trend of $f_{\rm b}$ evolutionary tracks. 
Figure~\ref{fig05} presents the mean $f_{\rm b}$ evolutionary track of the halo sample (with $M_{\rm halo} > 10^{11}\,\msol$ at $z=0.44$; purple line), together with the evolutionary track of a selected target halo (black line). The target halo broadly follows the mean trend, indicating that it is representative of the typical evolutionary behavior in the sample.

The gray dashed lines mark the two characteristic local extrema along the evolutionary tracks. After the peak, the baryon fraction decreases, whereas beyond the trough it increases again. Based on these features, the evolutionary history of halos can be divided into three distinct phases. 
\subsubsection{First phase: cold gas accretion }\label{sec:3result:2evol:0first}
During the first phase, corresponding to the increasing trend of the baryon fraction, dark matter collapses first, establishing the gravitational potential well, followed by the accretion of baryons into this pre-existing potential \citep{white1978}. 
The nearly monotonic increase in the baryon fraction reflects the fact that the halos in our sample form mainly at high redshift, when feedback processes are still weak or even absent. At these epochs, the gas is predominantly cold, allowing efficient baryon accretion with little thermal resistance.
\subsubsection{Second phase: SN-domination}\label{sec:3result:2evol:1SN}
Once the halos grow beyond a mass of $M_{\rm halo}\sim10^{10}\,\msol$, corresponding to the peak in the evolutionary tracks, they enter a second phase in which the baryon fraction begins to decline. This decrease can be understood as the combined outcome of two processes: (i) depletion of baryons from halos driven by strong outflows, and (ii) inefficient baryon accretion relative to dark matter growth, leading to a reduced baryon fraction.

Several physical mechanisms can, in principle, inhibit gas accretion, including mergers, virial shocks, SN feedback, and AGN feedback. However, the halos in this second phase are not massive enough to develop stable virial shocks \citep{virial_shock}. Moreover, in \NC, due to the adopted black hole seeding criteria, black holes are not typically seeded in halos with $M_{\rm halo} \lesssim 10^{11}\,\msol$. As a result, AGN feedback is largely absent in halos below this mass threshold. Even after a black hole is formed, AGN activity remains dominated by the quasar mode in halos with $M_{\rm halo}\lesssim10^{12}\,\msol$ in \NC\, where feedback primarily acts as isothermal heating that suppresses central star formation \citep{Teyssier2011} rather than driving large-scale gas outflows (jet mode). Consequently, AGN feedback is unlikely to play a major role in shaping the baryon fraction evolution during this phase.

This interpretation is further supported by the \textsc{NewHorizon2} simulation \citep{Yi2024}, which adopts almost identical subgrid feedback physics to \NC\ but suffers from unstable black hole settling problem that prevents many galaxies from retaining central black holes (see Figure~\ref{figA2}). Despite this, the $f_{\rm b}$ distributions for halos with $M_{\rm halo}\sim10^{10}$–$10^{12}\,\msol$ are remarkably similar between \textsc{NewHorizon2} and \NC\ (see Figure~\ref{figA} in Appendix~\ref{sec:app}). This close agreement supports the interpretation that the characteristic shape of the evolutionary tracks during the second phase is not primarily driven by AGN feedback, leaving SN feedback as the most plausible mechanism within the adopted \NC\ model.

\begin{figure}[htb!]
    \centering
    \includegraphics[width=0.35\textwidth]{\figpath/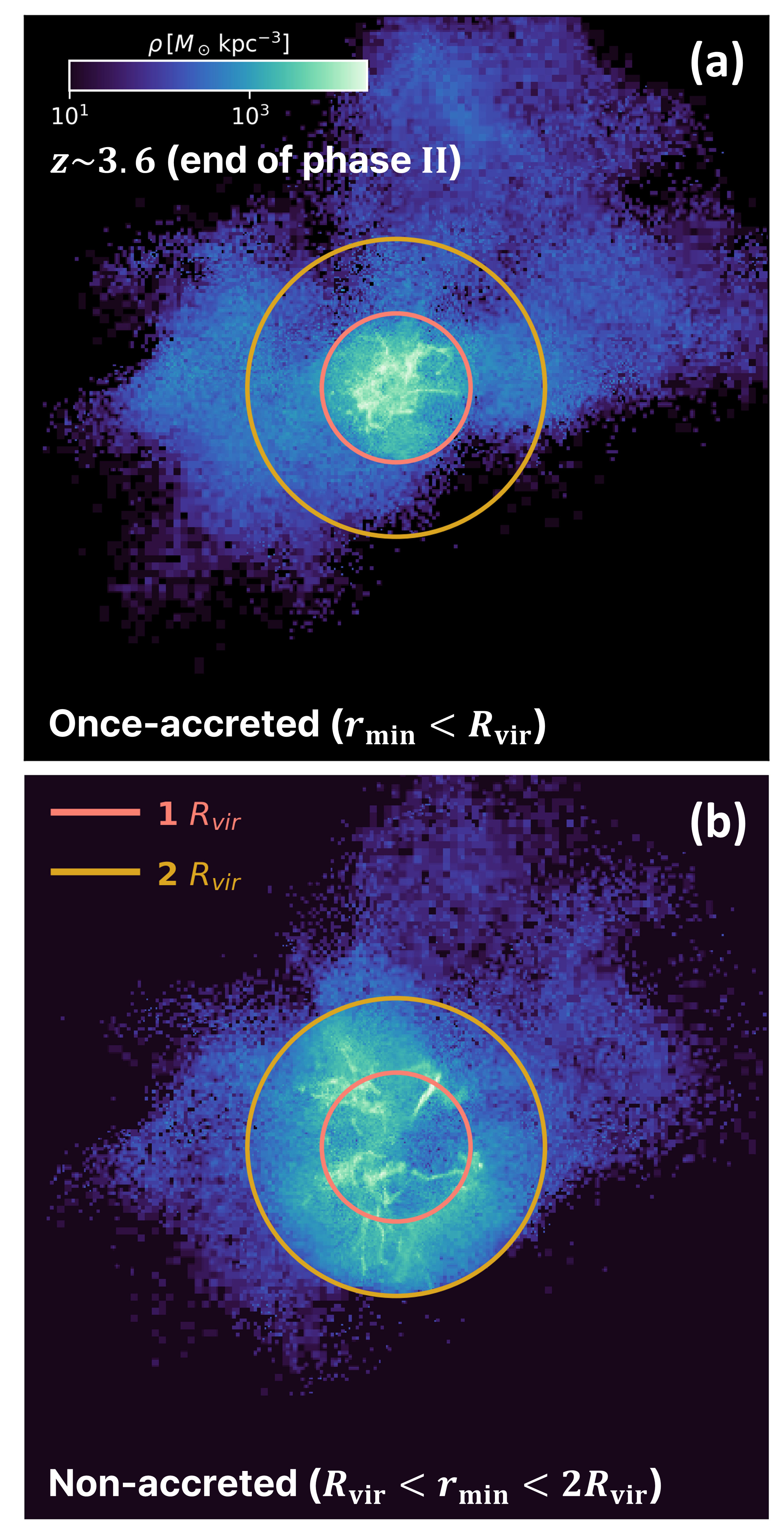}
    \caption{Spatial distribution of selected tracer particles associated with the target halo when it reaches the trough at $M_{\rm halo}\sim10^{12}\,\msol$. The tracers shown here are classified according to their radial histories with respect to the halo.
    Panel (a) presents tracer particles that have entered within $1\,R_{\rm vir}$ at least once. Tracers currently located outside $1\,R_{\rm vir}$ are therefore interpreted as outflowing gas. At this epoch, 18\% of these tracers have escaped beyond the virial radius.
    Panel (b) shows tracer particles that have remained exclusively within $1$–$2\,R_{\rm vir}$ at least once, but have never entered $1\,R_{\rm vir}$. Among these, 13\% are found beyond $2\,R_{\rm vir}$, indicating outward transport into the surrounding IGM.
    In both panels, the red and yellow circles denote $1\,R_{\rm vir}$ and $2\,R_{\rm vir}$, respectively.}
    \label{fig06}
\end{figure}

We utilize the tracer particles in order to investigate the role of outflows in $f_{\rm b}$ depletion, overcoming the limitation of the Eulerian approach. We select tracer particles associated with the target halo which is the same as the one in Figure~\ref{fig05}, based on their minimum radial distances. In Figure~\ref{fig06}, panel~(a) shows the spatial distribution of tracer particles at the end of the second phase, selected as those that resided within the halo virial radius at least once while the halo mass grew from $10^{10}$ to $10^{12}\,\msol$. A significant fraction of these tracers lie outside the virial radius (red circle), indicating significant influence of outflows. Quantitatively, approximately 18\% of the tracer particles escape beyond $R_{\rm vir}$ after reaching $r<R_{\rm vir}$.

However, as shown by the evolutionary track of the target halo in Figure~\ref{fig05}, the baryon fraction decreases from a value close to the cosmic $f_{\rm b}$ at the beginning of the second phase to roughly half of that value by its end. The removal of only 18\% of baryons by outflows is therefore insufficient to fully account for the observed decline in the baryon fraction, indicating that outflows alone cannot fully explain the suppression.

To capture additional processes that contribute to the reduction of baryon fraction, we present panel~(b). This panel shows tracer particles that satisfy two conditions: they have entered within $2\,R_{\rm vir}$ at least once during the second phase, but have never crossed within $R_{\rm vir}$. These tracers therefore represent gas residing exclusively in the halo’s surrounding environment. Their spatial distribution at the end of the second phase reveals that a non-negligible fraction of such surrounding gas is also displaced outward, with approximately 13\% of these tracers found beyond $2R_{\rm vir}$.

\begin{figure}[htb!]
    \includegraphics[width=0.45\textwidth]{\figpath/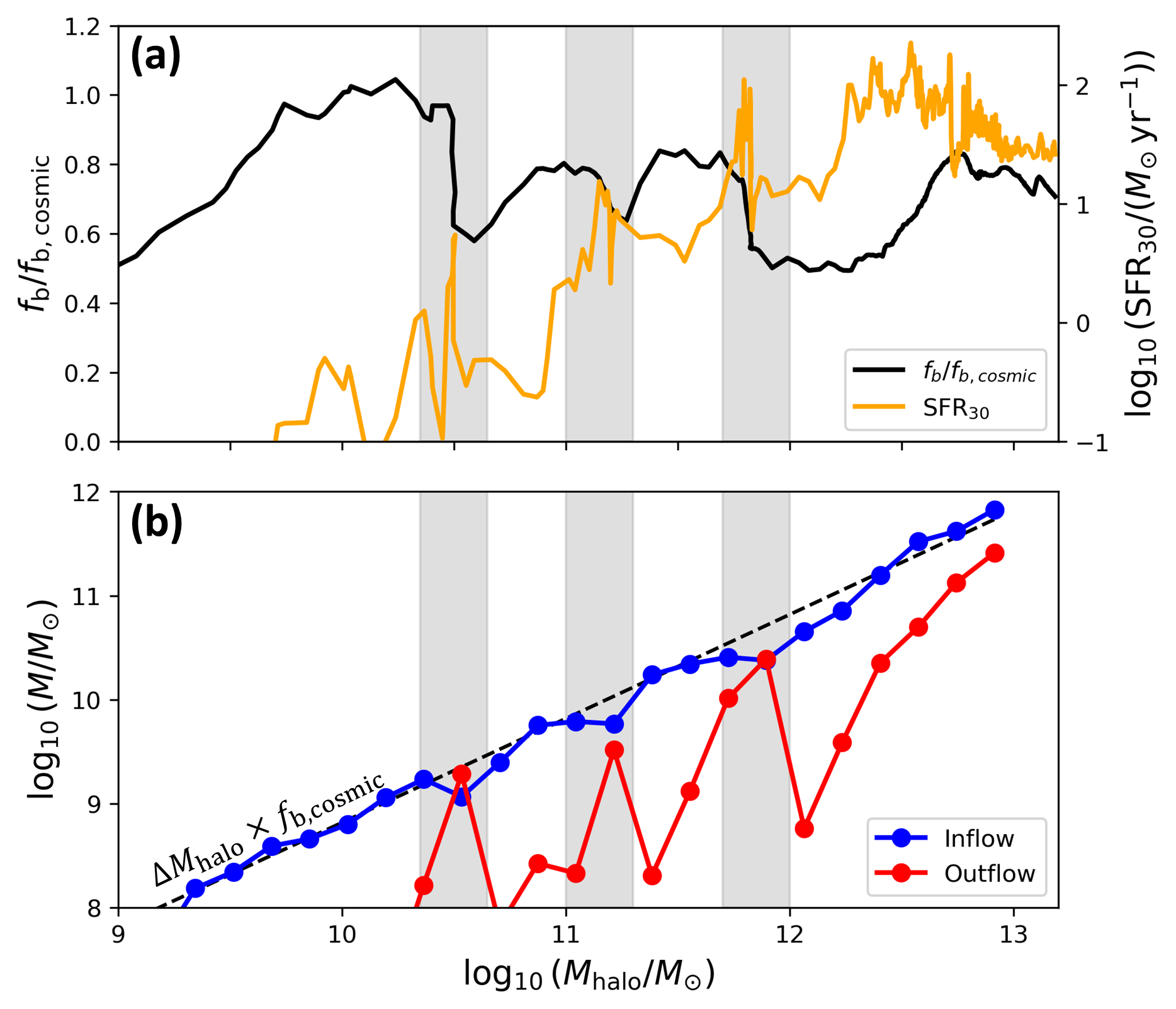}
    \caption{Coupled evolution of baryon fraction, star formation rate, and baryon flows in the target halo. In panel~(a), the black solid line represents $f_{\rm b}$ evolutionary track of the target halo, while the yellow line shows the star formation rate of the central galaxy averaged over the past 30 $\rm Myr$. The shaded gray regions indicate epochs where the baryon fraction exhibits a pronounced decline. During these epochs, enhanced star formation activity is observed, suggesting that baryon loss is accompanied by ongoing star formation.
    Panel~(b) presents the corresponding baryonic mass associated with inflow (blue) and outflow (red) as a function of halo mass, computed within discrete halo mass bins. The black dashed line represents $\Delta M_{\rm halo}\times f_{\rm b,cosmic}$, corresponding to the baryonic mass expected from halo mass growth assuming the cosmic baryon fraction. In the mass ranges where the baryon fraction decreases most rapidly, the outflow mass increases significantly, while the inflowing mass is reduced.}
    \label{fig07}
\end{figure}

To investigate the connection between these baryon-flow changes and star formation activity, as well as the role of preventative feedback \citep{Mitchell2021,Mitchell2022, Wright2024}, we analyze the coupled evolution of baryon fraction, star formation rate, and baryon inflow/outflow in the target halo (Figure~\ref{fig07}). Although the average evolutionary tracks in the second phase appear to decline monotonically, the individual target halo in fact experiences a more episodic evolution. While the overall trend is decreasing, we identify three distinct intervals during which the baryon fraction drops rapidly; these intervals are highlighted as gray shaded regions.

In panel~(a), the black line shows the evolutionary track of the selected target halo, previously introduced in Figure~\ref{fig05}, while the yellow line represents the star formation rate of the central galaxy averaged over the past 30\,$\rm Myr$. 
During the gray shaded intervals, the baryon fraction decreases while the star formation rate simultaneously increases, suggesting a close connection between baryon fraction suppression and star formation-driven feedback. In the first shaded region, a major merger event occurs. This merger triggers star formation, leading to baryon loss through feedback \citep{springel2005b,Hopkins2006}. In addition, while dark matter is relatively unaffected by stellar feedback and continues to accrete efficiently during the merger, baryons experience significant resistance \citep{Peirani2012}. As a result, the baryon fraction drops sharply, followed by a partial recovery as baryons accrete at later times. This behavior suggests that mergers can induce rapid, transient variations in the baryon fraction, but do not significantly alter the overall declining trend in the second phase. 

Panel~(b) presents the inflow (blue) and outflow (red) masses, inferred from tracer particles, measured at $R_{\rm vir}$ in the corresponding mass bins along the evolutionary track. 
During the gray shaded intervals, strong outflows are clearly observed, coinciding with the enhanced star formation activity seen in panel~(a). At the same time, inflow is suppressed relative to the baryonic mass expected from halo growth under the assumption of the cosmic baryon fraction, $\Delta M_{\rm halo}\times f_{\rm b,cosmic}$, during these periods. 
The combination of enhanced outflows and reduced inflows is consistent with a preventative-feedback scenario, in which feedback not only removes gas from within the halo but also inhibits the accretion of surrounding gas.
As also shown in panel~(b) of Figure~\ref{fig06}, this process can even expel gas residing in the surrounding environment.

Taken together, the tracer results directly demonstrate enhanced outflows and reduced inflows during the declining phase. The temporal association of these changes with enhanced star formation, together with the limited contribution of AGN feedback in this mass regime under the adopted black hole model, suggests that SN feedback is likely the dominant mechanism within \NC\ for expelling gas from halos and suppressing gas accretion. Our interpretation that SN feedback dominates in this mass range is consistent with previous studies based on statistical distributions at fixed redshift \citep{Mutch2013, Harrison2017, ay2023, Ema2025}, while our analysis provides a more direct evolutionary perspective by tracing halo assembly histories. We note, however, that the quantitative strength of the resulting baryon-fraction decrease likely depends on the adopted subgrid physics.
\subsubsection{Third phase: hot gas accretion}\label{sec:3result:2evol:2hot}
When the halo mass exceeds $\sim10^{12}\,\msol$, a third phase begins, in which $f_{\rm b}$ gradually recovers. This recovery implies an increase in the gas content relative to the dark matter mass, indicating a substantially improved gas retention capability in massive halos \citep{keres2005,virial_shock}.

During the preceding second phase, SN-driven feedback expels gas from halos and suppresses subsequent inflows. The affected gas is predominantly heated to high temperatures, making it difficult to remain gravitationally bound and thus causing it to reside beyond the virial radius. This behavior is evident in Figure~\ref{fig01}, panels (b) and (c), where the baryon fraction within the CGM ($R_{\rm gal} \lesssim r \lesssim R_{\rm vir}$) is lower than the cosmic value, while the surrounding IGM ($>R_{\rm vir}$) exhibits an excess of baryons. Such redistribution of baryons provides a natural explanation for the missing baryon problem observed in halos \citep{Bregman2007}.

Once halos grow to masses of $\gtrsim10^{12}\,\msol$, however, they become capable of sustaining hot-mode accretion. The deepened gravitational potential enables the accretion and retention of hot gas that was previously unable to cool or bind to the halo and thus remained in the IGM \citep{Kull1999, Penton2004, Eckert2015}. As a result, baryons that were heated and displaced by earlier feedback processes can be reincorporated into the halo, leading to the observed recovery of the baryon fraction during the third phase. This trend, in which increasingly massive halos approach the cosmic baryon fraction as their potential wells deepen toward group and cluster scales, is consistent with both previous numerical studies and observational constraints \citep{Gonzalez2013,Chiu2016,Chiu2018,ay2023,Ema2025}.
\begin{figure*}[htb!]
    \centering
    \includegraphics[width=0.95\textwidth]{\figpath/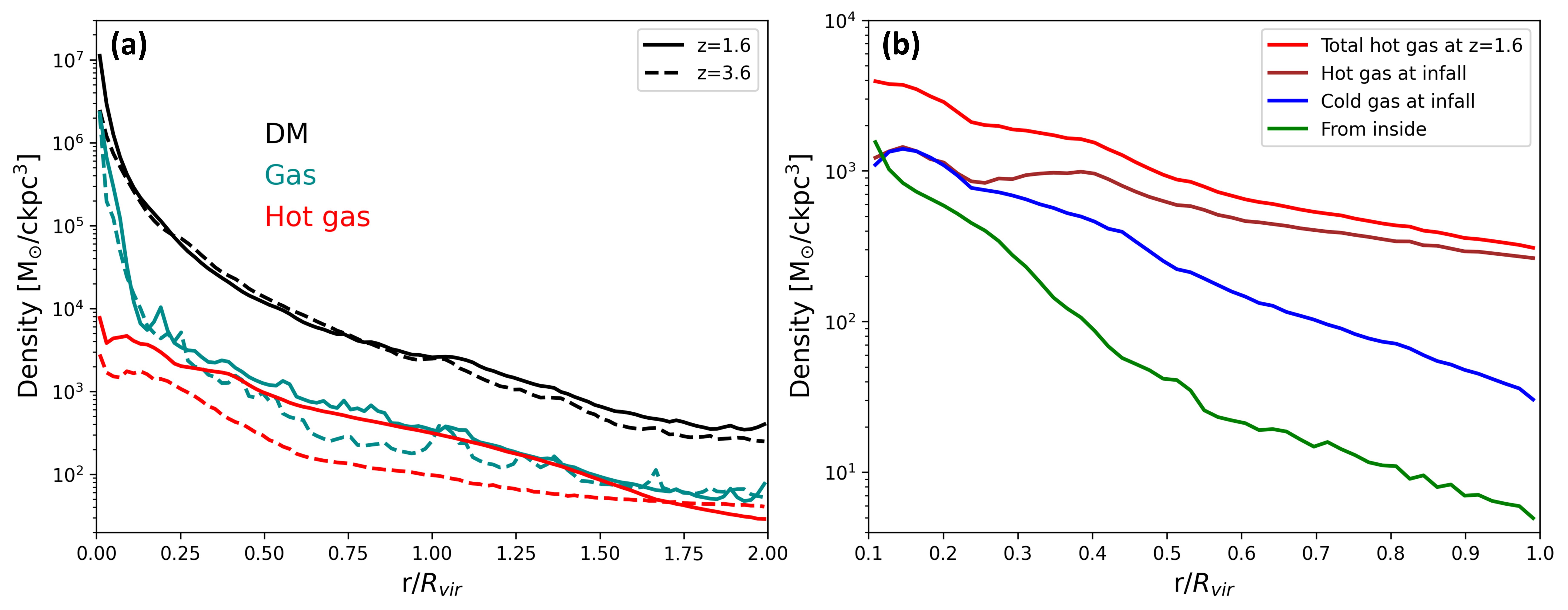}
    \caption{Radial density structure in a representative halo at two evolutionary stages and origin of hot gas at $z=1.6$.
    Panel (a) compares the radial density profiles of dark matter (black), total gas (dark cyan), and hot gas (red) for a halo at two evolutionary stages: when the halo is located at the trough ($M_{\rm halo}\sim10^{12}\,\msol$ at $z=3.6$, dashed lines) and when the baryon fraction increases again at later times ($M_{\rm halo}\sim5\times10^{12}\,\msol$ at $z=1.6$, solid lines). Distances are normalized by the virial radius. While the shapes of dark matter profiles remain similar, the gas component becomes more enriched in the CGM at $z=1.6$, and the hot gas density increases significantly over the entire radial range.
    Panel (b) shows the radial density profile of hot gas at $z=1.6$, classified by its origin using tracer particles: gas expelled from the central galaxy (green), gas that was already hot at the time of infall (brown), and gas that was cold at infall (blue). The hot gas component is dominated by gas that was already hot prior to accretion.}
    \label{fig08}
\end{figure*}

To investigate the evolution underlying this picture, we compare the radial density profiles of dark matter, gas, and hot gas ($T\ge10^{6}\,\mathrm{K}$) for the target halo shown in Figure~\ref{fig05} at two representative epochs, as summarized in Figure~\ref{fig08}. Panel~(a) compares the radial density profiles of dark matter, gas, and hot gas for the target halo at two epochs: at the end of the second phase, when the halo mass reaches $M_{\rm halo}\simeq10^{12}\,\msol$ ($z=3.6$, dashed lines), and at a later time when the baryon fraction has been substantially recovered, $M_{\rm halo}\simeq5\times10^{12}\,\msol$ ($z=1.6$, solid lines).

The dark matter density profiles show little difference between the two stages when expressed in units of $R_{\rm vir}$, indicating that the underlying gravitational potential structure remains largely unchanged. In contrast, the gas density profile exhibits a noticeable enhancement in the circumgalactic region, approximately $0.25\lesssim r/R_{\rm vir}\lesssim1$, implying that the increase in the baryon fraction is primarily driven by the accumulation of gas in the CGM.

Moreover, the hot gas component shows a significant increase across nearly the entire radial range, suggesting that the halo becomes increasingly efficient at retaining and accreting hot gas. This supports the interpretation that the recovery of the baryon fraction in the third phase is closely linked to the onset of efficient hot-mode accretion in massive halos.

However, the increase in hot gas within the CGM does not by itself prove that hot-mode accretion is the only source. 
It could also arise from feedback processes originating in the central galaxy. Disentangling these two effects is therefore essential.
We decompose the hot gas profile at $z=1.6$ according to its origin and address in panel~(b).
The hot gas is classified into three categories: (i) gas originating from the central region (“from inside”), (ii) accreted gas that was already hot at the time of infall ($T\ge10^{6}\,\mathrm{K}$), and (iii) accreted gas that was cold at infall ($T<10^{6}\,\mathrm{K}$). Here, inflowing gas is defined using tracer particles that entered the halo after $z=3.6$. Tracers that were already present at $z=3.6$ and do not belong to category (i) are not included in this classification; however, they constitute only about 2\% of the total hot gas and are therefore negligible.

To identify hot gas expelled by internal feedback, the galactic region is defined as the inner $0.1\,R_{\rm vir}$  \citep[e.g.,][]{Kocjan2024}. Tracer particles that passed through this region at least once are classified as “from inside.” For the remaining tracers, the gas temperature at the moment they crossed $R_{\rm vir}$ is used to distinguish between hot and cold infall.
Hot gas that was already hot at infall is interpreted as evidence for the halo’s ability to accrete hot gas owing to its deepened potential well, whereas hot gas that entered cold and subsequently heated, as well as gas originating from the inner region, is attributed to feedback-related processes and virial heating in sufficiently massive halos.

As shown in panel~(b), the increase in hot gas density in the radial range $0.25\lesssim r/R_{\rm vir}\lesssim1$ is dominated by gas that was already hot at infall. This provides direct evidence that \emph{the enhanced hot gas content in the CGM originates predominantly from hot-mode accretion, rather than from in-situ heating by feedback processes.}

In this mass regime ($M_{\rm halo}>10^{12}\,\msol$), most galaxies host central black holes with properties broadly consistent with observational constraints \citep{newcluster}.
Also, AGN jet mode feedback is activated in most halos of this mass range and redshift~\citep{Dubois2012,Volonteri2016,Habouzit2022}. Since the onset of the jet mode depends primarily on the Eddington ratio rather than halo mass itself, the characteristic halo mass at which the jet mode is triggered varies from halo to halo. Once activated, the jet mode can drive strong outflows, leading to a temporary reduction in the halo baryon fraction. In fact, such a decrease was visible in Figure~\ref{fig05} at $M_{\rm halo} \gtrsim 5\times10^{12}\,\msol$. However, because the activation mass and timing differ among individual halos, these episodic decreases are largely averaged out in the evolutionary tracks and do not appear as a coherent trend. Nevertheless, without sufficiently effective AGN feedback, the increase of the baryon fraction during the third phase becomes steeper, as seen for \textsc{NewHorizon2} in Figure~\ref{figA}.


Therefore, in the third phase, the recovery of the baryon fraction is primarily driven by the halo’s increased mass and deepened potential well, which enable the accretion of gas that was already hot at infall. Our results show that this externally supplied hot gas dominates the CGM, establishing hot-mode accretion as the primary origin of the recovered baryon content. While AGN jet-mode feedback can induce decreases in the baryon fraction, these effects are transient and subdominant.
\section{Discussion}\label{sec:4discussion}

\subsection{Comparison with observations on galaxy scale}\label{sec:4discussion:1obs}

As direct observational constraints on halo-scale baryon fractions remain limited, especially in the low-mass regime, we instead compare galaxy-scale properties of our simulation with observations, where measurements are more robust. While agreement at the galaxy scale does not guarantee consistency at the halo scale, it provides a meaningful observational baseline for assessing the credibility of the baryonic physics realized in the simulation in the absence of direct halo-scale constraints.

To this end, we compare our \NC\ sample with observational measurements from the MAGPI (Middle Ages Galaxy Properties with Integral Field Spectroscopy) survey \citep{MAGPI2021}. MAGPI is an integral-field spectroscopic survey conducted with MUSE on the VLT, providing spatially resolved kinematics of stars and ionized gas for galaxies at intermediate redshift, with a median redshift of $\langle z \rangle \sim 0.35$, corresponding to look-back times of $\sim$3--4\,Gyr.

Using the MAGPI data, \citet{Sharma2026} constructed a sample of star-forming galaxies with stellar masses of $10^{8}\mbox{--}10^{11.5}\,\msol$, comprising 397 galaxies over the redshift range $z = 0.1\mbox{--}0.85$, and measured the dark matter fraction on galaxy scales.

Dark matter fraction $f_{\rm DM}$ within the effective radius was estimated by decomposing the dynamical mass as
\begin{equation}
M_{\rm dyn} \!=\! M_{\rm disk} \!+\! M_{\rm bulge} \!+\! 1.33\,M_{\rm HI} \!+\! M_{\rm H_2} \!+\! M_{\rm DM},
\end{equation}
where $M_{\rm disk} + M_{\rm bulge}$ were derived from spectral energy distribution fitting, $M_{\rm HI}$ and $M_{\rm H_2}$ gas masses were inferred using scaling relations, and $M_{\rm dyn}$ was obtained from rotation curve fitting. The factor of 1.33 in HI gas accounts for the contribution of helium to the atomic gas mass. For further methodological details, we refer the reader to \citet{Sharma2026}.

For a fair comparison with MAGPI, we construct a galaxy sample from the last snapshot of \NC\ by applying selection criteria as closely matched as possible to the observations. 
The redshift of the last snapshot of \NC\ (0.44) is reasonably close to the median redshift of the MAGPI data.
Since satellite galaxies are strongly affected by environmental processes, we restrict our analysis to central galaxies only. We further apply the same stellar mass cut as in the MAGPI sample, selecting galaxies with $10^{8} \le M_{\star}/\msol \le 10^{11.5}$. In addition, the MAGPI analysis accounts only for cold gas when estimating the baryonic mass. To ensure consistency, we therefore consider only gas with temperatures below $10^{4}\,$K, corresponding to the HI temperature regime, and densities above $1\,{\rm H\,cm^{-3}}$ in the \NC\ sample. Based on these criteria, we measure the stellar mass and cold gas mass within the half-mass radius and define the baryonic mass as
$M_{\rm baryon} = M_\star + M_{\rm cold\,gas}$.

In the approach adopted by \citet{Sharma2026}, the dark matter mass is inferred by subtracting the stellar and cold gas components from the dynamical mass, such that warm or hot gas components are not explicitly removed, even though they are implicitly included in the dynamical mass estimate. To apply the same limitation consistently to \NC, we compute this pseudo dark matter mass by subtracting $M_{\rm baryon}$ from the total matter mass within the half-mass radius.

\begin{figure}[tb!]
    \includegraphics[width=0.45\textwidth]{\figpath/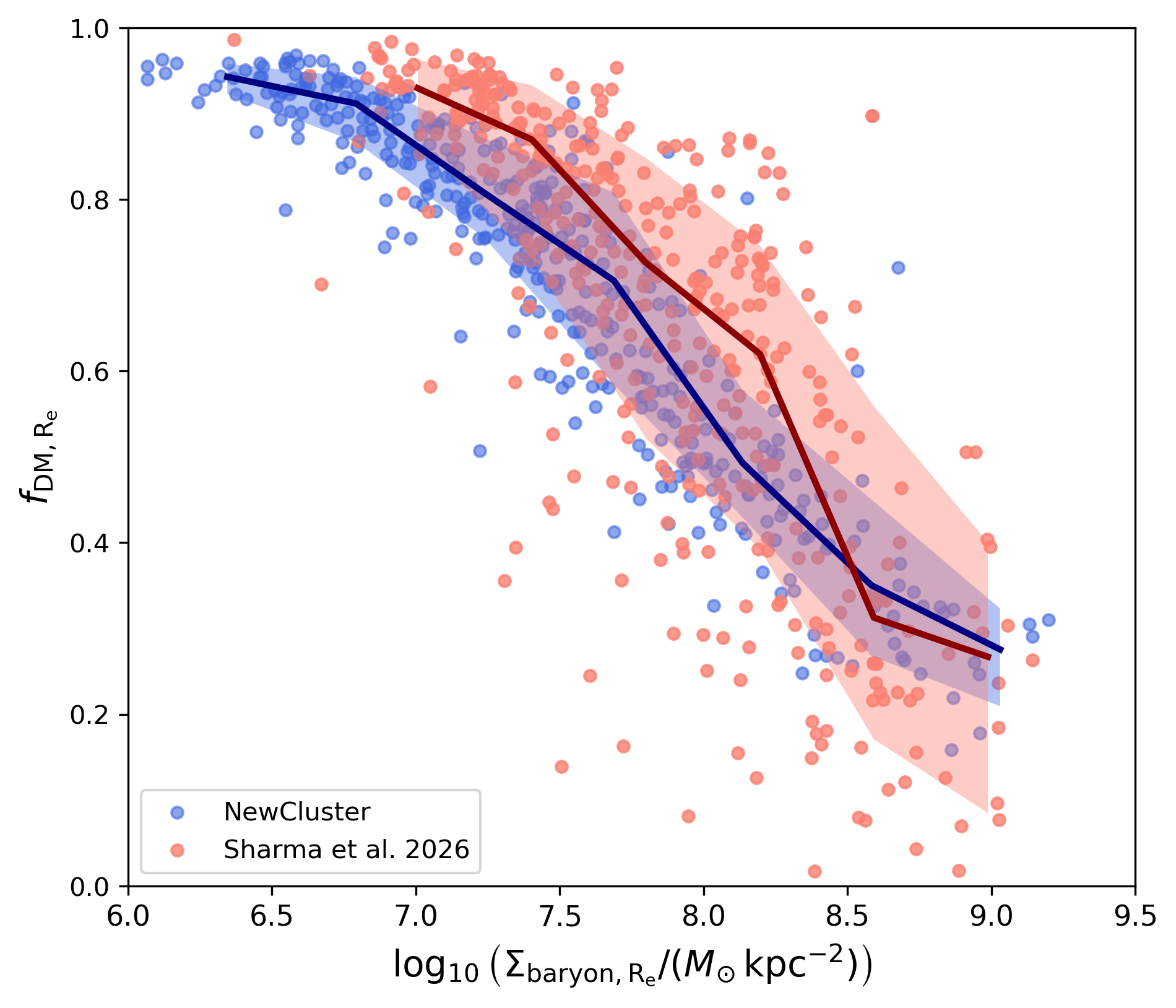}
    \caption{Comparison of galaxy-scale $f_{\rm DM}$ between \NC\ at $z=0.44$ and observations. $f_{\rm DM}$ within the effective radius is shown as a function of the baryonic surface density, $\Sigma_{\rm baryon}$, measured within the same radius (or the half-mass radius for \NC). Blue points represent central galaxies from \NC, while red points correspond to the MAGPI sample from \citet{Sharma2026}. The solid lines indicate the median trends for each sample, and the shaded regions represent the 16th--84th percentile ranges.}
    \label{fig09}
\end{figure}

Figure~\ref{fig09} presents a comparison between MAGPI and \NC. The $x$-axis shows the baryonic surface density within the effective radius (or the half-mass radius for \NC), while the $y$-axis indicates $f_{\rm DM}$ within the corresponding region. We find that the \NC\ galaxies broadly match the observed galaxy-scale $f_{\rm DM}$ measured in the MAGPI sample. Although $f_{\rm b}$ in the CGM is known to be highly sensitive to subgrid feedback prescriptions, galaxy-scale baryon and dark matter fractions are also directly affected by these processes. The agreement with MAGPI therefore provides an important observational consistency check on the galaxy-scale baryonic physics realized in \NC\ and a meaningful basis for investigating the baryon cycle on larger scales. We emphasize, however, that agreement on galaxy scales does not by itself guarantee an accurate baryon distribution at the halo scale.

\subsection{Environmental effects}\label{sec:4discussion:2env}
\begin{figure}[tb!]
    \includegraphics[width=0.45\textwidth]{\figpath/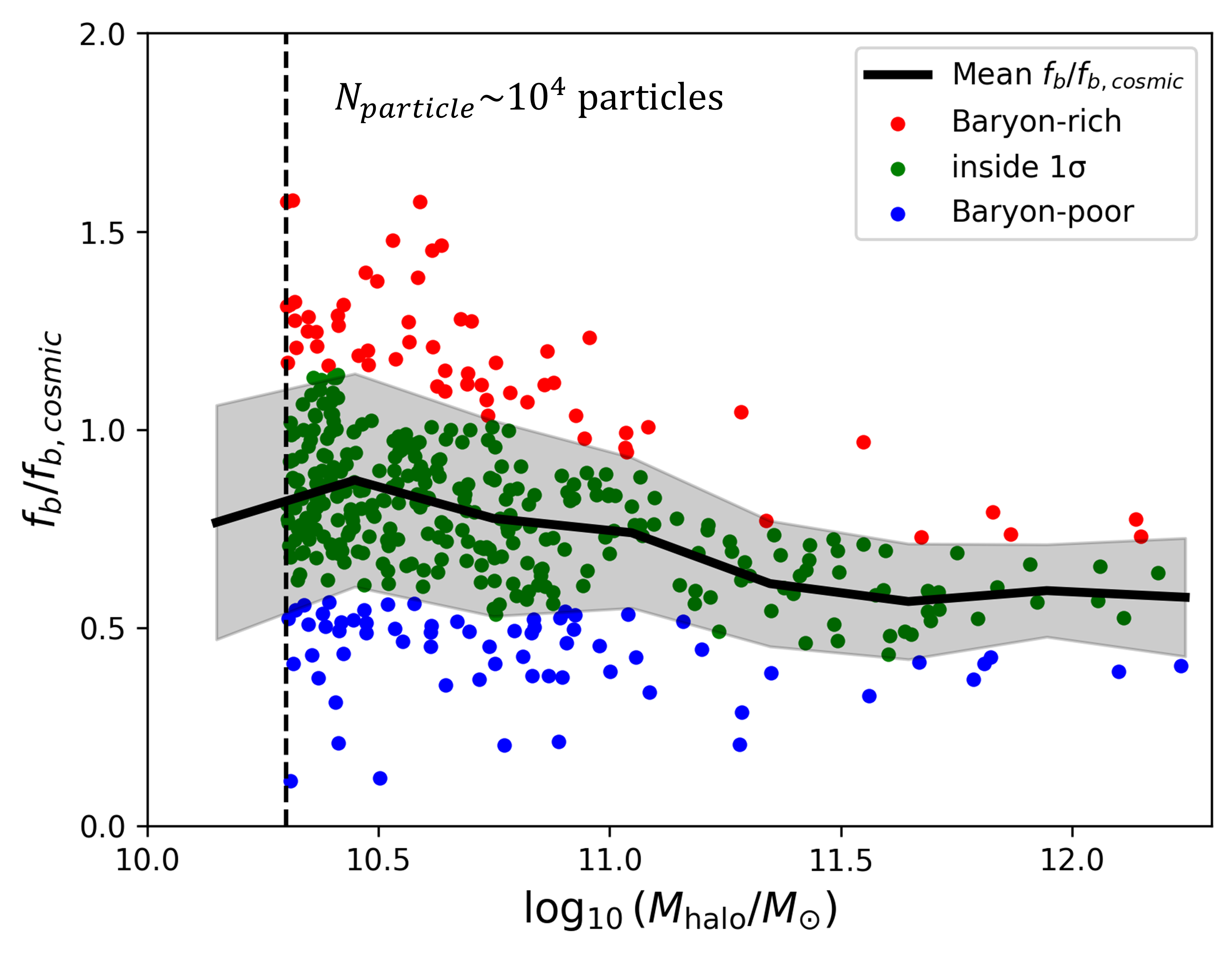}
    \caption{Halo baryon fraction, normalized by the cosmic value, as a function of halo mass for main halos in \NC\ at $z=0.8$. The black solid line shows the mean $f_{\rm b}/f_{\rm b,cosmic}$ at fixed mass, and the gray shaded region indicates the $1\sigma$ scatter. Halos are classified based on their deviation from the mean: baryon-rich (BR; red), within $1\sigma$ (green), and baryon-poor (BP; blue). Only halos containing more than $\sim 10^4$ dark matter particles are included in the analysis, corresponding to the mass threshold indicated by the vertical dashed line. The substantial dispersion at fixed halo mass motivates an investigation of environmental effects beyond internal feedback processes.}
    \label{fig10}
\end{figure}
\begin{figure*}[htb!]
    \centering
    \includegraphics[width=0.95\textwidth]{\figpath/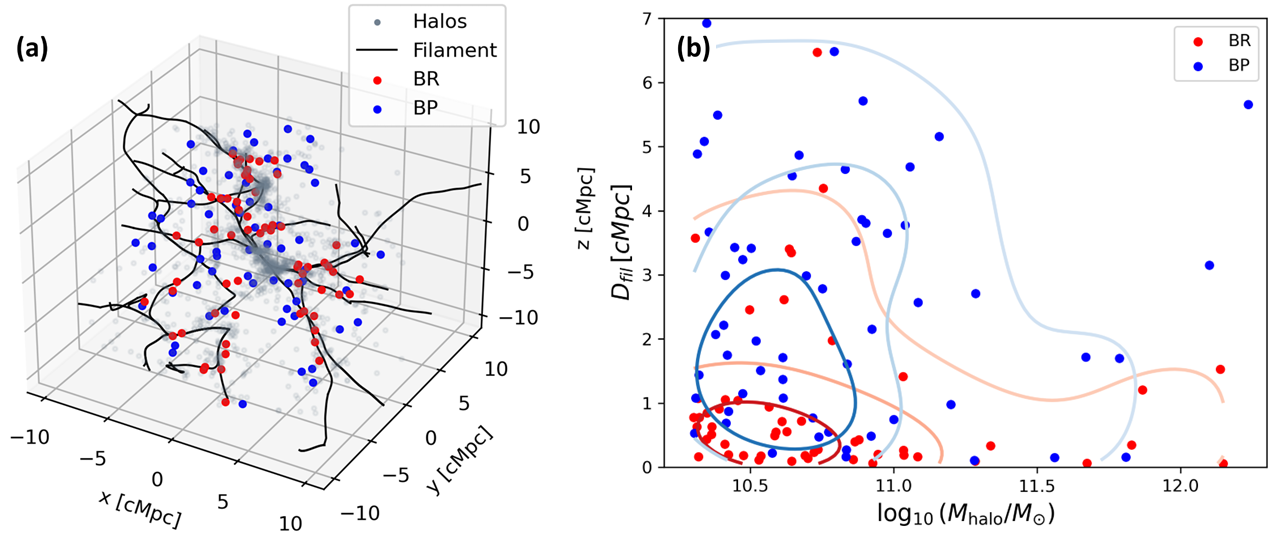}
    \caption{
        Spatial correlation between halo baryon fraction and proximity to cosmic filaments at $z=0.8$.
        Panel~(a) shows the spatial distribution of halos and filaments. The filamentary structures, identified from the dark matter density field constructed using the cloud-in-cell (CIC) method with additional smoothing and extracted via DisPerSE, are shown as black solid lines. Individual halos are represented by gray dots, with baryon-rich (BR) halos marked in red and baryon-poor (BP) halos in blue.
        Panel~(b) presents the distribution of the shortest perpendicular distances from each BR and BP halo to the nearest filament, based on the spatial configuration shown in panel~(a). BR halos tend to have systematically smaller distances to filaments compared to BP halos. Contours are computed from a two-dimensional (2D) kernel density estimate and correspond to regions enclosing 25\%, 50\%, and 90\% of the data.}
    \label{fig11}
\end{figure*}

While the overall shape of the baryon fraction evolutionary tracks appears to be primarily governed by internal processes and surrounding gas conditions, a significant halo-to-halo scatter is present at fixed halo mass. In particular, halos with similar masses can exhibit substantially different baryon fractions during their evolution.
\cite{musso18} used excursion theory to explain how the geometry and connectivity of the cosmic web influence halo assembly and the statistics of matter accretion. 
Their study may provide the origin for the large scatter in $f_{\rm b}$ we found.

Here, we focus on $z = 0.8$, corresponding to the epoch just before cluster-scale mergers become dynamically important in \NC. This choice allows us to minimize the impact of merger-driven interactions and better isolate the role of the filamentary cosmic web environment in shaping the scatter of baryon fraction.

To quantify this diversity, we classify halos according to their deviation from the mean $f_{\rm b}$ at a fixed mass. Figure~\ref{fig10} shows the distribution of $f_{\rm b}$, where halos with baryon fractions greater than $1\sigma$ above the mean are defined as ``baryon-rich'' (hereafter BR; 62 halos), those within $1\sigma$ as inside $1\sigma$, and those more than $1\sigma$ below the mean as ``baryon-poor'' (hereafter BP; 66 halos). To ensure sufficient numerical resolution, we restrict our analysis to halos containing more than $\sim 10^4$ dark matter particles, corresponding to a mass cut of $M_{\rm halo} \gtrsim 2\times10^{10}\,\msol$. Halos with $M_{\rm halo} \gtrsim 2\times10^{12}\,\msol$ are excluded due to their limited sample size, which prevents a robust separation into BR and BP.

Even within a narrow halo mass range, the BR and BP samples can differ in baryonic mass by more than a factor of two. Such a large dispersion suggests that in addition to internal processes and surrounding gas, more external factors may also contribute to shaping the baryon content of halos.

Since our sample consists exclusively of main halos rather than satellites embedded in groups or clusters, strong local environmental processes such as ram-pressure stripping \citep[e.g.,][]{Kenny2004, Chung2009, Jung2018, jaffe2018, Rhee2026} are unlikely to be the dominant driver of the observed scatter. It is therefore more reasonable to investigate the influence of the large-scale cosmic environment, particularly the connection to filamentary structures of the cosmic web \citep{Tuominen2021}. Halos located closer to filaments may experience different gas supply, thermal conditions, and dynamical interactions
\citep[e.g.,][]{Bond1996,keres2005,Pichon2011,Codis2012,Kraljic2018}.

To characterize the cosmic web structure, we employ the publicly available Discrete Persistent Structure Extractor (DisPerSE; \citealt{Sousbie2011a, Sousbie2011b}). DisPerSE is a topological structure-finding algorithm designed to identify multiscale features in a density field.
Indeed, \cite{musso18} showed that topology is the correct metric for understanding halo growth. They showed how the connectivity and local curvature of the cosmic web (as treated in an extended excursion theory framework) dictate the `Lagrangian history' of galactic assembly.
This approach enables a robust identification of filamentary structures across a wide range of scales.

To extract the filamentary structure, we first construct a density field within the central region of \NC, corresponding to a comoving volume of $(42.6\,{\rm cMpc})^3$. The density field is generated on a uniform grid with a resolution of approximately $166\,{\rm ckpc}$ using the cloud-in-cell (CIC) interpolation scheme. To reduce small-scale noise and obtain a smoother representation of the large-scale structure, we further apply an additional smoothing procedure to the gridded density field.

DisPerSE is then applied to the resulting smoothed density field. We adopt a cut value of 0.7, which corresponds approximately to a persistence ratio equivalent to a $4\sigma$ significance level. This threshold effectively suppresses spurious structures arising from sampling noise while retaining the dominant filamentary network, and corresponds to a regime where the large-scale structural distribution becomes well separated from noise-induced features.

In Figure~\ref{fig11}, panel~(a) shows the filamentary structure identified by DisPerSE. Gray dots represent the spatial distribution of all halos, while red and blue symbols indicate the BR and BP halos, respectively. For each halo, we measure the shortest perpendicular distance to the nearest filament, and the resulting distance distributions are presented in panel~(b). In this analysis, we exclude halos located within $3\,R_{\rm vir}$ ($4.26\,{\rm cMpc}$) of the main cluster, since the environmental effects of the massive cluster cannot be neglected in such regions. 
After this selection, 52 BR and 57 BP halos remain.

In panel~(b), it is clear that BR halos are found closer to filaments, compared to BP halos.
These results demonstrate a correlation between halo baryon fraction and proximity to filamentary structures. A two-sample Kolmogorov--Smirnov (KS) test performed on the $D_{\rm fil}$ distributions in panel~(b) confirms that the BR and BP halo populations are statistically distinct at high significance, yielding $D_{\rm KS}=0.525$ and $p=2.1\times10^{-7}$.

However, one important limitation of the DisPerSE-based filament identification is that the extracted filament structure does not provide information on filament width. As a result, halos located at the same perpendicular distance from a filament may reside in physically different environments depending on the filament density structure. To account for this effect, we additionally measured cylindrical radial dark matter density profiles around individual filament segments and estimated the overdensity associated with each halo based on its radial position relative to nearby filaments. We note that the resulting overdensity trends are broadly consistent with those inferred from filament distance alone.

These results indicate a correlation between halo baryon fraction and the global filamentary environment. We note, however, that proximity to filaments does not necessarily imply more efficient gas accretion, since both gas density and temperature increase toward filament spines. In the mass range considered here ($M_{\rm halo}\gtrsim2\times10^{10}\,\msol$), where redshift-dependent baryon-fraction suppression is very weak, BR halos reside closer to filaments despite their relatively hot surroundings. In an additional test (plot not shown) extending to lower halo masses, where suppression becomes stronger, the BR--BP difference in filament proximity weakens, while BP halos tend to reside in hotter IGM environments, suggesting an increasing importance of the thermal state of the surrounding gas.

We therefore interpret the present result as an environmental correlation rather than direct evidence that filament proximity enhances gas accretion. Establishing the physical connection between filamentary environment and baryon acquisition requires a detailed analysis of individual gas accretion histories, which is beyond the scope of the present study and will be addressed in future work.

\subsection{Caveats}\label{sec:5caveat}

One of the key uncertainties in this work arises from the adopted subgrid physics, which plays a central role in regulating the baryon cycle. As discussed in the introduction paper of \NC\ \citep{newcluster}, the simulation exhibits a tendency to overestimate stellar masses in low-mass halos, particularly around $M_{\rm halo}\sim10^{11}\,\msol$, as indicated by the stellar-to-halo mass relation \citep{Moster2010, Kravtsov2018, Behroozi2019}. Elevated stellar masses can lead to stronger SN feedback, which may in turn affect the predicted halo baryon fraction in this mass regime. This suggests that the quantitative strength of the baryon-fraction decrease during the second phase likely depends on the adopted feedback model.

In addition, the black hole seeding model in \NC\ adopts stringent criteria, resulting in black holes being rarely seeded in halos with $M_{\rm halo}\lesssim10^{11}\,\msol$. Consequently, the impact of black hole feedback cannot be assessed in this lower-mass regime.
However, some of these limitations are not expected to qualitatively alter our main conclusions. 
Although black holes are not efficiently seeded below $10^{11}\,\msol$, feedback from AGN is unlikely to play an important role in regulating the baryon fraction in such low-mass halos.

In addition, since \NC\ is a zoom-in simulation, the statistical sample size is inherently smaller than those of typical full-volume cosmological simulations, which may lead to reduced statistical robustness and render cosmic variance non-negligible. 
Moreover, because \NC\ follows the formation of a massive cluster environment, it may not fully represent the average environment of the Universe.
Despite that, as presented in Appendix~\ref{sec:app}, the difference between \NC\ and \textsc{NewHorizon2} in terms of the baryon fraction trend is not large.
Our additional comparison with the full-volume \textsc{Horizon-AGN} simulation, also presented in Appendix~\ref{sec:app}, shows that low-mass halos in cluster environments have somewhat lower baryon fractions than those in non-cluster environments, while the overall trend remains similar.


\section{Conclusions}\label{sec:6conclusion}
In this work, we investigate the evolution of the baryon fraction within dark matter halos using the \NC\ cosmological zoom-in simulation, with the aim of linking baryon fraction evolution to the underlying physical processes governing halo growth. Rather than simply connecting evolutionary pathways solely from comparisons of $f_{\rm b}$ distributions at different redshifts, we directly trace individual halo histories using merger trees and tracer particles, thereby connecting baryon fraction evolution to the baryon cycle. 

The high mass and spatial resolution of \NC, combined with its short snapshot intervals, enables a detailed investigation across cosmic time, while the use of tracer particles allows us to track gas motions and origins, making \NC\ particularly well suited for both statistical characterization of the $f_{\rm b}$ distribution and reconstruction of individual evolutionary tracks.

We trace the evolutionary histories of 65 main halos with $M_{\rm halo} > 10^{11}\,\msol$ at $z = 0.44$. Our analysis reveals an $N$-shaped trend in which the baryon fraction evolution proceeds through three distinct phases. 
\begin{enumerate}
    \item First phase: it corresponds to the early cosmic times, during which efficient cold accretion allows relatively cold gas to be readily incorporated into assembling dark matter halos. 
    \item Second phase: the baryon fraction appears to decline primarily due to SN feedback. Tracer analysis directly reveals enhanced outflows and reduced subsequent inflow, indicating a significant role of preventative feedback, while their association with enhanced star formation supports SN feedback as the dominant mechanism in this regime.
    \item Third phase: in more massive halos, we identify a distinct recovery phase, in which deeper gravitational potentials enable efficient gas retention and reaccretion, driving the baryon fraction back toward the cosmic value. Tracer analysis reveals that this recovery is dominated by gas that was already hot at infall, demonstrating that hot-mode accretion plays a key role in restoring the baryon content. This highlights a clear transition in the dominant baryon cycle and represents a key result of this work.
\end{enumerate}

We find that the $f_{\rm b}$ distribution at fixed redshift does not unambiguously encode the evolutionary pathways of halos. Halos with similar masses and baryon fractions at a given epoch can follow markedly different assembly histories, particularly in the low-mass regime where strong redshift dependence is observed. This suggests that population-level statistics alone are insufficient to capture the time-dependent nature of the baryon cycle. In particular, a direct comparison between the $f_{\rm b}$ distribution and individual evolutionary tracks reveals a systematic shift in the location of the first peak, implying that the distribution alone may misrepresent the underlying physical trends.

We further find that baryon fraction suppression in low-mass halos appears to be linked to the thermal state of the surrounding IGM. In particular, strong suppression emerges after the IGM temperature becomes higher than the halo virial temperature for an extended period of time, highlighting the critical role of environmental heating in regulating baryon accretion. In addition, the scatter in baryon fraction at fixed halo mass is correlated with the large-scale environment, with baryon-rich halos tending to reside closer to filamentary structures in mass regimes where thermal suppression remains weak.

Overall, our results demonstrate that baryon fraction evolution is governed by a complex interplay between internal feedback processes and environmental effects, and cannot be fully understood without sufficient spatial resolution and a time-resolved perspective on halo growth. These findings have important implications for the interpretation of baryon fraction measurements and for comparisons between simulations. Given that the halo baryon fraction is very sensitive to the adopted subgrid models, continued improvements and systematic comparisons across different simulations are essential to robustly uncover the physical origin of baryon distributions in the Universe.

\section*{Acknowledgements}
We thank the referee for providing comments that helped clarify several points in the early manuscript. Special thanks go to Gauri Sharma for providing the MAGPI observational data in Figure~\ref{fig09}. 
This work was granted access to the HPC resources of KISTI under the allocations KSC-2021-CRE-0486, KSC-2022-CRE-0088, KSC-2022-CRE-0344, KSC-2022-CRE-0409, KSC-2023-CRE-0343, KSC-2024-CHA-0009, and KSC-2025-CRE-0031 and of GENCI under the allocations A0150414625 and A0180416216.
The large data transfer was supported by KREONET, which is managed and operated by KISTI.
S.K.Y. acknowledges support from the Korean National Research Foundation (RS-2025-00514475 and RS-2022-NR070872).
J.R. acknowledges support from the Institut de Physique des deux infinis of Sorbonne Universit\'{e} and by the ANR grant ANR-19-CE31-0017 of the French Agence Nationale de la Recherche.
E.C. and C.S. acknowledge support from the Korean National Research Foundation (RS-2023-00241934).
S.J. acknowledges support from the Korea Astronomy and Space Science Institute under the R\&D program (Project No. 2026-1-831-03) supervised by the Ministry of Science and ICT.


\bibliography{reference}{}
\bibliographystyle{aasjournal}
\clearpage
\appendix
\section{baryon fraction distributions in \textsc{RAMSES} simulations}\label{sec:app}
\renewcommand{\thefigure}{A\arabic{figure}}
\setcounter{figure}{0}
\begin{figure}[htb!]
    \includegraphics[width=0.45\textwidth]{\figpath/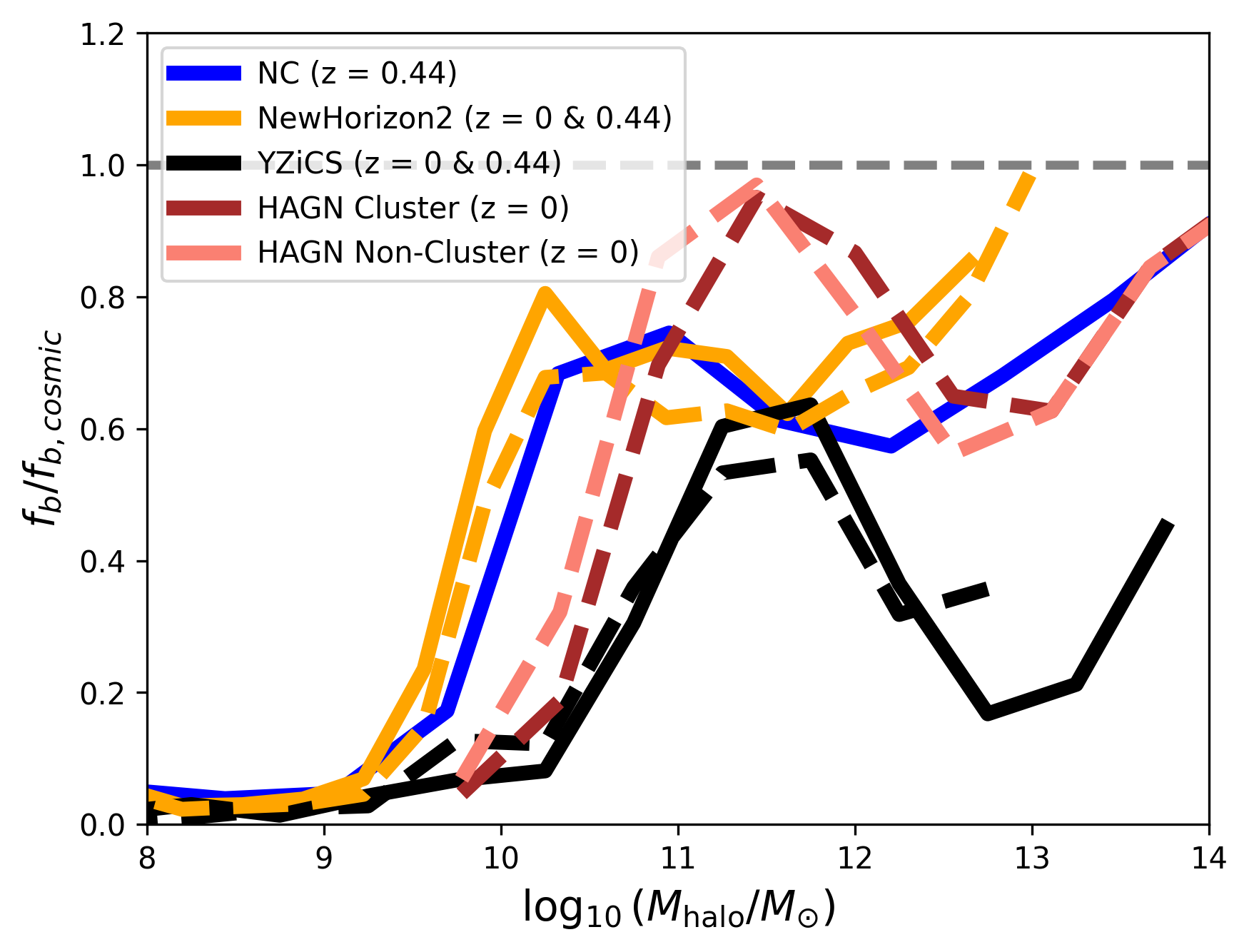}
    \caption{
    $f_{\rm b}$ distributions in \textsc{RAMSES} simulations.
    The blue, orange, and black curves represent the zoom-in simulations
    \NC, \textsc{NewHorizon2}, and \textsc{YZiCS}, respectively, while the light- and dark-red curves represent halos from the full-volume \textsc{Horizon-AGN} simulation, divided into cluster and non-cluster environments, respectively.
    Solid and dashed curves indicate $z=0.44$ and $z\sim0$, respectively.
    }
    \label{figA}
\end{figure}
In this appendix, we compare the $f_{\rm b}$ distributions of several \textsc{RAMSES}-based simulations. Figure~\ref{figA} shows the $f_{\rm b}$ distributions of \NC\ (blue), \textsc{NewHorizon2} (orange), and \textsc{YZiCS} (black) at $z=0.44$ (solid) and $z\sim0$ (dashed).

For \textsc{YZiCS} \citep{YZiCS2017}, the distribution becomes similar
to that of \textsc{SIMBA}, suggesting that the distinct distribution in
\NC\ is not simply a consequence of the zoom-in setup. Both
\textsc{YZiCS} and \textsc{NewHorizon2} show little evolution in
$f_{\rm b}$ between $z=0.44$ and $z=0$, indicating that the
\NC\ distribution at $z=0.44$ is broadly representative of the
low-redshift regime.

To further assess the possible impact of the overdense environment sampled by \NC, we additionally use the full-volume \textsc{Horizon-AGN} simulation to compare halos in cluster and non-cluster environments at $z\sim0$, defining the former as regions within 14 Mpc of 12 halos with $M_{\rm vir}>10^{14}\,\msol$, approximately matching the characteristic radial extent of the \NC\ zoom-in region. As shown in Figure~\ref{figA}, low-mass halos in cluster environments tend to have somewhat lower baryon fractions than their counterparts in non-cluster environments up to $M_{\rm halo} \sim 10^{11} \msol  $, while the overall trend remains similar.

This environmental suppression is qualitatively consistent with the slightly lower baryon fractions of \NC\ relative to \textsc{NewHorizon2} at low halo masses. However, the magnitude of the environmental suppression is very weak, indicating that the large-scale environment is unlikely to be the primary origin of the overall trend.

\begin{figure}[htb!]
    \includegraphics[width=0.45\textwidth]{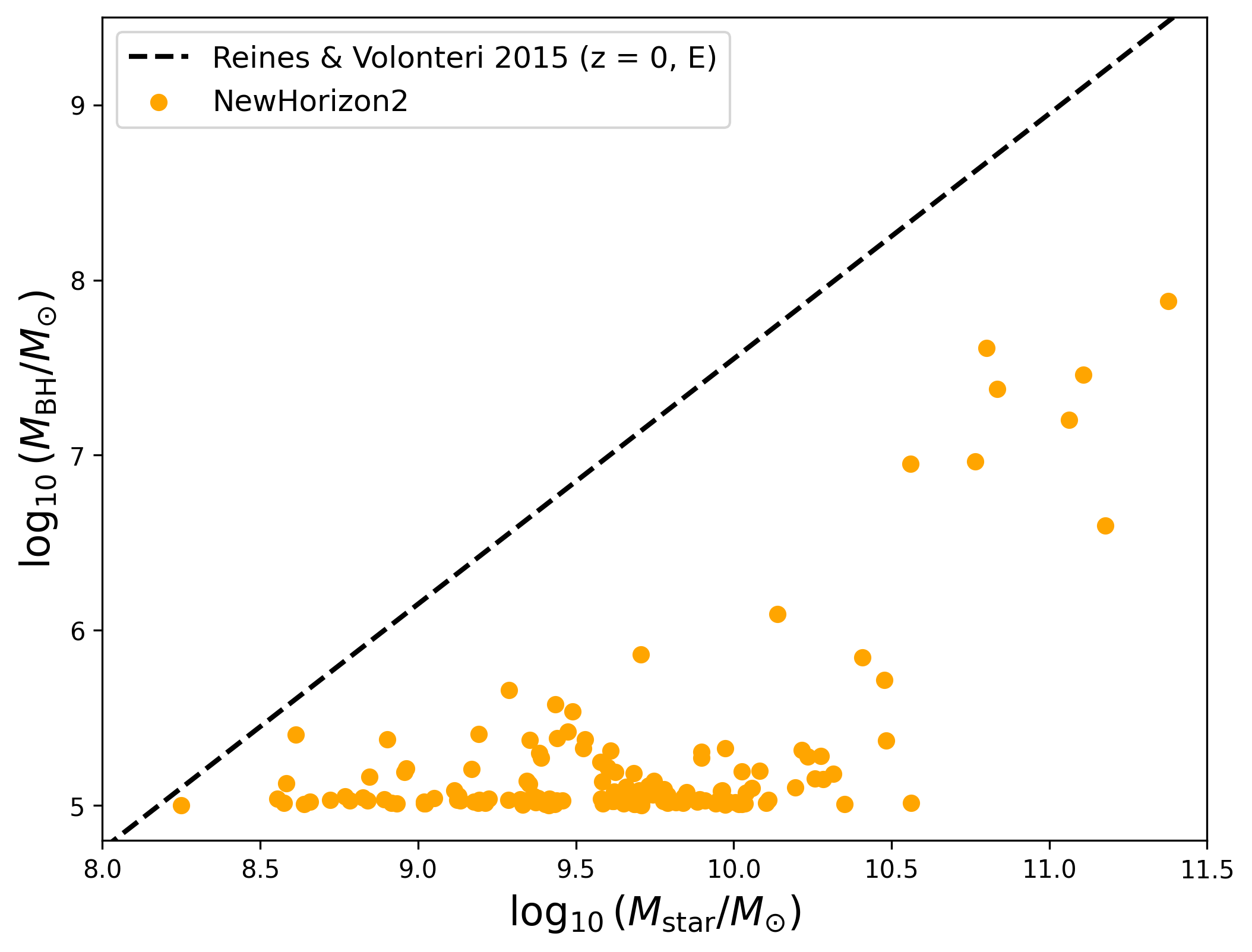}
    \caption{Black hole mass versus stellar mass for the \textsc{NewHorizon2} galaxies (orange circles). The black dashed line shows the $z=0$ empirical relation from \citet{Reines2015}.}
    \label{figA2}
\end{figure}

The offset in the peak between \textsc{YZiCS} and \NC\ at $z=0.44$ is likely driven not only by differences in subgrid physics, but also by differences in the dark matter mass resolution. In the case of \textsc{NewHorizon2}, the distribution at $z=0.44$ closely follows that of \NC. In particular, the similarity in the mass range $10^{10}$--$10^{12}\,\msol$, despite the absence of well-centered black holes and their systematically undermassive nature (see Figure~\ref{figA2}), indicates that AGN feedback is not the dominant process in this regime. Also, since black holes in \textsc{NewHorizon2} are not well centered within their host halos, AGN feedback cannot operate efficiently. This likely causes the third evolutionary phase to begin at slightly lower halo masses and leads to weaker suppression of gas inflow, resulting in a steeper increase in baryon fraction with halo mass.

\end{document}